\documentclass[11pt,onecolumn,amssymb,nofootinbib]{revtex4}
\usepackage{amsmath, amsthm, amscd, amssymb}
\usepackage{bm}
\usepackage{bbm}

\begin{document}

\title{\bf Classical and Quantum Gauged Massless Rarita-Schwinger fields}

\author{Stephen L. Adler}
\email{adler@ias.edu} \affiliation{Institute for Advanced Study,
Einstein Drive, Princeton, NJ 08540, USA.}

\begin{abstract}
We show that, in contrast to known results in the massive case, a minimally gauged
massless Rarita-Schwinger field yields consistent classical and quantum theories, with a
generalized fermionic gauge invariance.
To simplify the algebra, we study a two-component left chiral reduction of the massless theory.
We formulate the classical theory in both Lagrangian and Hamiltonian form for a general
non-Abelian gauging, and analyze the constraints and the Rarita-Schwinger gauge invariance of the action.
An explicit wave front calculation for
Abelian gauge fields shows that wave-like modes do not propagate with superluminal velocities.
The quantized case is studied in gauge covariant radiation gauge and $\Psi_0=0$ gauge  for the Rarita-Schwinger field, by both functional integral and Dirac bracket methods. In $\Psi_0=0$ gauge, the constraints have the form needed to apply the Faddeev-Popov method for deriving a
functional integral. The Dirac bracket approach in $\Psi_0=0$ gauge yields consistent Hamilton equations of motion, and in covariant radiation
gauge leads to anticommutation relations with the correct positivity properties. We discuss relativistic covariance of the
 anticommutation relations, and of Rarita-Schwinger scattering from an Abelian potential.  We note that fermionic gauge transformations are a canonical transformation, but further details of the transformation between different fermionic gauges are left as an open problem.

\end{abstract}

\maketitle

\section{Introduction}

\subsection{Motivations and Background}
Cancelation of gauge anomalies is a basic requirement for constructing grand unified models, and the usual
assumption is that anomalies must cancel among spin $\frac{1}{2}$ fermion fields.  However,  a 1985 paper
of Marcus \cite{marcus} showed that in principle an $SU(8)$ gauge theory can be constructed with
spin $\frac{3}{2}$ Rarita-Schwinger fermions playing a role in anomaly cancelation, and we have recently
constructed \cite{adler} a family unification model incorporating this observation.  Using gauged spin $\frac{3}{2}$ fields in a grand unification model raises the question of whether such fields admit a consistent quantum, or even classical
theory. It is well known, from papers of Johnson and Sudarshan \cite{johnson} and Velo and Zwanziger \cite{velo}, and
much subsequent literature (see e.g. Deser and Waldron \cite{deser}),  that theories of massive gauged Rarita-Schwinger fields have serious problems. Does setting
the fermion mass to zero eliminate these difficulties?

The lesson we have learned from the success of the Standard Model is that fundamental fermion masses lead to
problems and are to be avoided; all mass is generated by spontaneous symmetry breaking, either through
coupling to the Higgs boson or through the formation of  chiral symmetry breaking fermion condensates.  So from a modern point
of view, the Rarita-Schwinger theory with an explicit mass term is suspect. Several  hints that the behavior of the massless
theory may be satisfactory are already apparent from a study of the zero mass limit of formulas in the Velo--Zwanziger paper.  First, in their demonstration of superluminal
signaling, the problematic sign change that they find for large $B$ fields \big(Eq. (2.15) of \cite{velo}\big) is not present when the mass is set to zero.
Second, when the mass is zero, the secondary constraint that they derive \big(Eq. (2.10) of \cite{velo}\big) appears as a factor in the change in the action  under a Rarita-Schwinger gauging $\delta \psi_{\mu} = D_{\mu} \epsilon$, with $D_{\mu}$ the usual gauge covariant derivative. \big(This statement is not in \cite{velo}, but is an easy calculation from their Eqs. (2.1)--(2.3), with the $D_{\mu}$ of this paper their $-i\pi_{\mu}$.\big)  Hence, the constrained action in the massless gauged Rarita-Schwinger  theory has a fermionic gauge invariance that is the natural generalization of the fermionic
gauge invariance of the massless free Rarita-Schwinger theory.  Third, their formula for the anticommutator \big(Eq. (4.12) of
\cite{velo}\big)  in the zero mass case
develops an apparent singularity  in the limit of vanishing gauge field $B$, and so their quantization does not limit to the standard free
theory quantization.  This should not be surprising: since the massive theory does not have a fermionic gauge invariance, Ref.
\cite{velo} does not include a gauge-fixing term analogous to that used in the massless case, but gauge fixing is needed
to get a consistent quantum theory for a free massless Rarita-Schwinger field.  So these observations, following from the
equations in \cite{velo}, suggest that a study of the massless Rarita-Schwinger field coupled to spin-1 gauge fields is in order.

In a different setting, massless Rarita-Schwinger fields appear consistently coupled to gravity as the gravitinos of supergravity, as discussed by
Das and Freedman \cite{dasa}.  Grisaru, Pendleton  and van Nieuwenhuizen \cite{grisaru}  have shown that soft spin $\frac{3}{2}$ fermions {\it must} be coupled to
gravity as in supergravity, in an analysis based on the free particle external line pole structure of spin $\frac{3}{2}$ fields that do not have spin 1 gauge couplings. Their result does not rule out the possibility of gauged spin $\frac{3}{2}$ fields with
non-Abelian gauge couplings, since such fermion fields will in general not have free particle external lines.  For example, a gravitino coupled in the fundamental representation of an unbroken non-Abelian gauge group  will be confined, and will not have the kind of external line pole structure used in the argument of  \cite{grisaru}.  Thus, the known connection of ungauged
Rarita-Schwinger fields to supergravity does not argue against the possibility that there could be a consistent theory of
massless, gauged Rarita-Schwinger fields, so again a detailed study of this possibility is warranted.

\subsection{Outline of the paper, and summary}
With these motivations and background in mind, we embark in this paper on a detailed study of the classical and quantum theory of a minimally gauged massless Rarita-Schwinger field.  In Sec. 2, we give the Lorentz covariant Lagrangian for a gauged four-component Rarita-Schwinger spinor field, derive the source current for the gauge field, and check that it is gauge-covariantly conserved.  We also give the Lorentz covariant form of the constraints and of the fermionic gauge invariance, and of the symmetric stress-energy tensor. In Sec. 3 we argue that the massless, gauged
 Rarita-Schwinger theory  exhibits a generalized fermionic gauge invariance, in the sense that the action principle does not uniquely
 determine the system time evolution.  Since in the massless
case left chiral and right chiral components of the field decouple,  in Sec. 4 we rewrite the Lagrangian for left
chiral components in terms of two-component spinors and Pauli matrices, which simplifies the subsequent analysis.  We then give
the Euler-Lagrange equations in two-component form, and use them to analyze the structure of constraints and the fermionic gauge invariance of the action.  In Sec. 5 we specialize to the case of an Abelian gauge field \big(as in \cite{velo}\big) and analyze the wavefront structure, showing that wave modes  propagate with
subluminal velocities; an extension of this discussion is given in Appendix B.    In Sec. 6 we return to the general case of non-Abelian gauge fields. We introduce canonical momenta for the Rarita-Schwinger field components, which are used to define classical Poisson brackets,  and discuss the role of the constraints as generators of gauge transformations under the bracket operation.  We  show that the constraints group into two sets of four, within each of which there are vanishing Poisson brackets.
This permits application of the Faddeev--Popov method for  path integral quantization, which we carry through in detail in Sec. 7 in $\Psi_0=0$ gauge.
In Sec. 8  we give the Hamiltonian form of the
equations of motion and constraints, and introduce the Dirac bracket.  When a gauge fixing constraint is omitted, the Dirac bracket that we compute agrees with the anticommutator calculated in \cite{johnson} and \cite{velo}.  In Sec. 9  we study the Dirac bracket in its classical and quantum forms, and show in covariant radiation gauge that the quantum Dirac bracket has the requisite positivity
properties to be an anticommutator; related details are given in Appendix C.  In Sec. 10 we study the behavior  under Lorentz transformations of covariant radiation
gauge, of  $\Psi_0=0$ gauge,  and of the Dirac bracket and anticommutator. In Sec. 11 we analyze Rarita-Schwinger fermion scattering from a
 short range Abelian gauge potential, using an analog of the distorted wave Born approimation.  We show consistency of relativistic
 covariance with helicity state counting, but there is no small-coupling expansion giving an unmodified Born approximation for this process. In Sec. 12 we discuss areas for extensions of our results, and in Appendix A we summarize our notational conventions and some useful identities.  We suggest that the reader skim through Appendix  A before going on to Sec. 2, since things stated in Appendix A are not repeated in the body of the paper.

Our conclusion from this analysis is that one can consistently gauge a massless Rarita-Schwinger field, at both the classical
and quantum levels.   This opens the possibility of using gauged Rarita-Schwinger fields as part of the anomaly cancelation mechanism in grand unified models, with anomalies of the spin $\frac{1}{2}$ fields canceling against the
spin $\frac{3}{2}$ anomaly.

\section{Lagrangian and covariant current conservation in four-component form}

The action for the massless Rarita-Schwinger theory is
\begin{align}\label{eq:action}
S(\psi_{\mu}) = &\frac{1}{2} \int d^4x \,\overline{\psi}_{\mu \alpha u} R^{\mu \alpha u}~~~,\cr
 R^{\mu\alpha u}=&i\epsilon^{\mu\eta\nu\rho}(\gamma_5\gamma_{\eta})^{\alpha}_{~\beta}(D_{\nu}\psi_{\rho}^{\beta})^u~~~,\cr
 (D_{\nu}\psi_{\rho}^{\beta})^u\equiv&\partial_{\nu}\psi_{\rho}^{\beta u}+gA_{\nu v}^{u} \psi_{\rho}^{\beta v}~~~,\cr
 A_{\nu v}^u=&A_{\nu}^A t_{A v}^u~~~,\cr
\end{align}
with $\psi^{\mu \alpha u}$ a four-vector four-component spinor, with four-vector index $\mu=0,...,3$, spinor index $\alpha=1,...,4$, and $SU(n)$ internal
symmetry index $u=1,...,n$, with $SU(n)$ gauge generators $t_A,\,A=1,...,n^2-1$.  Taking  $u$ to range from 1 to  $n$ means that, for definiteness, we are
 assuming that the spinors transform according to the fundamental representation of the $SU(n)$ internal symmetry group, but other representations and other compact Lie groups can be
 accommodated by assigning the  internal indices $u$ and $A$ the appropriate range.  Note that $t_A$, $ A_{\nu v}^u$, and $D_{\nu}$ all commute with the gamma matrices and  the Pauli
spin matrices from which the gamma matrices are constructed, and for an Abelian internal symmetry group, the indices $u$ and $A$ are not needed.  Using
\begin{equation}
 \overline{\psi}_{\mu\alpha u}=\psi_{\mu\beta u}^{\dagger}i(\gamma^0)^{\beta}_{~\alpha}~~~,
\end{equation}
 together with the adjoint convention $(\chi_1^{\dagger}\chi_2)^{\dagger}=\chi_2^{\dagger}\chi_1$ for Grassmann variables
 $\chi_1,\,\chi_2$, it is easy to verify that $S$ is self-adjoint.

From here on we will usually not indicate the spinor indices $\alpha,\, \beta$ and internal symmetry indices $u,\,v$ explicitly, but they are implicit in all
formulas.   Varying $S$ with respect to the
Rarita-Schwinger fields, we get the equations of motion
\begin{align}\label{eq:eqmo}
\epsilon^{\mu\eta\nu\rho} \partial_{\nu}\overline{\psi}_{\rho}\gamma_{\eta}=& g\epsilon^{\mu\eta\nu\rho}
\overline{\psi}_{\rho}A_{\nu}^A t_{A} \gamma_{\eta}~~~,\cr
\epsilon^{\mu\eta\nu\rho}\gamma_{\eta}\partial_{\nu}\psi_{\rho}=&-g \epsilon^{\mu\eta\nu\rho}
\gamma_{\eta} A_{\nu}^A t_{A} \psi_{\rho}~~~.\cr
\end{align}
Re-expressed in terms of the covariant derivative, these are
\begin{align}\label{eq:eqmo1a}
\epsilon^{\mu\eta\nu\rho} \overline{\psi}_{\rho}\overleftarrow{D}_{\nu}\gamma_{\eta}=&0~~~,\cr
\epsilon^{\mu\eta\nu\rho}\gamma_{\eta}D_{\nu}\psi_{\rho}=&0~~~.\cr
\end{align}

The $\mu=0$ component of these equations gives the primary constraints
\begin{align}\label{eq:constraint1}
\epsilon^{enr} \overline{\psi}_{r}\overleftarrow{D}_{n}\gamma_{e}=&0~~~,\cr
\epsilon^{enr} \gamma_{e}D_{n}\psi_{r}=&0~~~,\cr
\end{align}
with $e,n,r$ summed from 1 to 3.
Contracting the equation of motion for $\overline{\psi}_{\rho}$  with $g^{-1}\overleftarrow{D}_{\mu}$ and
the equation of motion for $\psi_{\rho}$ with $g^{-1}D_{\mu}$, we get the secondary constraints
\begin{align}\label{eq:constraint2}
\epsilon^{\mu\eta\nu\rho} \overline{\psi}_{\rho}F_{\mu\nu}\gamma_{\eta}=&0~~~,\cr
\epsilon^{\mu\eta\nu\rho}\gamma_{\eta}F_{\mu\nu}\psi_{\rho}=&0~~~,\cr
\end{align}
 where we have introduced the  gauge field strength
\begin{align}\label{eq:gaugefield}
F_{\mu\nu}=&g^{-1}[D_{\mu},D_{\nu}]=g^{-1}[\overleftarrow{D}_{\mu},\overleftarrow{D}_{\nu}]\cr
=&\partial_{\mu}A_{\nu}-\partial_{\nu}A_{\mu}+g[A_{\mu},A_{\nu}]~~~,\cr
\end{align}
which with the adjoint representation index $A$ indicated explicitly reads
\begin{equation}\label{eq:gaugefield1}
F_{\mu\nu}^A=\partial_{\mu}A_{\nu}^A-\partial_{\nu}A_{\mu}^A+g f_{ABC}A_{\mu}^B A_{\nu}^C~~~.
\end{equation}
Under a Rarita-Schwinger gauge transformation (with $\epsilon$ a four-component spinor), which is
a natural gauge field generalization of the fermionic gauge invariance for a free, massless Rarita-Schwinger field discussed
in \cite{dasb},
\begin{align}\label{eq:gaugetrans}
\psi_{\mu} \to&  \psi_{\mu}+ \delta_G \psi_{\mu}~~,~~~   \delta_G \psi_{\mu}   \equiv      D_{\mu}\epsilon~~~,\cr
\overline{\psi}_{\mu} \to &  \overline{\psi}_{\mu}+ \delta_G  \overline{\psi}_{\mu} ~~,~~~ \delta_G  \overline{\psi}_{\mu} \equiv  \overline{\epsilon} \overleftarrow{D}_{\mu}~~~,\cr
\end{align}
the action of Eq. \eqref{eq:action} changes according to
\begin{equation}\label{eq:actionchange}
\delta_G S(\psi_{\mu})= -\frac{1}{4}ig \int d^4x \Big[ \overline{\epsilon} \gamma_5 \Big(\epsilon^{\mu\eta\nu\rho} \gamma_{\eta} F_{\mu\nu} \psi_{\rho}\Big)
+\Big(\epsilon^{\mu\eta\nu\rho} \overline{\psi}_{\rho} F_{\mu\nu} \gamma_{\eta}\Big) \gamma_5 \epsilon\Big] ~~~.
\end{equation}
The factors bracketed in large parentheses are identical to the secondary constraints of Eq. \eqref{eq:constraint2}.  We will argue in the next
section that this implies that, even when coupled to gauge fields,  the Rarita-Schwinger theory has a generalized form of fermionic gauge invariance.

Adding the gauge field action
\begin{equation}\label{eq:gaugeaction}
S(A_{\mu}^A)=-\frac{1}{4}\int d^4x F_{\mu\nu}^A F^{A\mu\nu}~~~,
\end{equation}
and varying the sum $S(\psi_{\mu})+S(A_{\mu}^A)$ with respect to the gauge potential, we get the
gauge field equation of motion
\begin{align}\label{eq:gaugemo}
D_{\nu}F^{A\mu\nu}&\equiv\partial_{\nu}F^{A\mu\nu}+g f_{ABC} A_{\nu}^B F^{C\mu\nu}=gJ^{A\mu}~~~,\cr
J^{A\mu}=&\frac{1}{2}\overline{\psi}_{\nu}i\epsilon^{\nu\eta\mu\rho}\gamma_5\gamma_{\eta}t_{A} \psi_{\rho}~~~.\cr
\end{align}
A straightforward calculation using Eqs. \eqref{eq:eqmo} shows that the gauge field source current $J^{A\mu}$ obeys the
covariant conservation equation
\begin{equation}\label{eq:jcons}
D_{\mu}J^{A\mu}=\partial_{\mu}J^{A\mu}+gf_{ABC}A_{\mu}^B J^{C \mu}=0~~~,
\end{equation}
as required for consistency of Eq. \eqref{eq:gaugemo}.
So from the Rarita-Schwinger and gauge field actions, we have obtained a formally consistent set of equations of motion.

In addition to the gauge field source current, there is an additional current $J^{\mu}$ that obeys an ordinary conservation
equation,
\begin{align}\label{eq:fermion}
J^{\mu}=&\frac{1}{2}\overline{\psi}_{\nu}\epsilon^{\nu\eta\mu\rho}\gamma_5\gamma_{\eta}\psi_{\rho}~~~,\cr
\partial_{\mu}J^{\mu}=&0~~~.\cr
\end{align}
In the massive spinor case, Velo and Zwanziger \cite{velo} argue that the analogous current, within the constraint subspace of Eq. \eqref{eq:constraint1}, should have a positive time component. In  the massless case we see no reason for this requirement, since
Eq. \eqref{eq:fermion} is the fermion number current, and its time component, giving the fermion number density, can have either sign. However, we shall use parts of the positivity
argument of \cite{velo} later on in discussing positivity of the Dirac bracket anticommutator.

The symmetric stress-energy tensor for the free massless Rarita-Schwinger has been computed by Das \cite{das1} (see also Allcock and Hall \cite{allcock}).   Changing ordinary derivatives to
gauge covariant derivatives, Das's formula becomes
\begin{align}\label{eq:rstensor}
T_{\rm RS}^{\sigma\tau}=&-\frac{i}{4}\epsilon^{\lambda\mu\nu\rho}\Big[\overline{\psi}_{\lambda} \gamma_5 (\gamma^{\tau}\delta^{\sigma}_{\mu}+ \gamma^{\sigma} \delta^{\tau}_{\mu}) D_{\nu}\psi_{\rho}\cr
+&\frac{1}{4}\partial_{\alpha}\Big(\overline{\psi}_{\lambda} \gamma_5 \gamma_{\mu}([\gamma^{\alpha},\gamma^{\sigma}]\delta^{\tau}_{\nu}
+ [\gamma^{\alpha},\gamma^{\tau}]\delta^{\sigma}_{\nu})\psi_{\rho}\Big)\Big]~~~.\cr
\end{align}
\big(This formula can be made manifestly self-adjoint by replacing $D_{\nu}$ by $\frac{1}{2}(D_{\nu}-\overleftarrow{D}_{\nu})$, but this is not needed to verify
stress-energy tensor conservation.\big)  Adding the gauge field stress-energy tensor,
\begin{equation}\label{eq:gauge}
T_{\rm gauge}^{\sigma\tau}=-\frac{1}{4}\eta^{\sigma\tau}F^A_{\lambda\mu}F^{A\lambda\mu} + F^{A\sigma}_{\lambda}F^{A\lambda\tau}~~~,
\end{equation}
a lengthy calculation, using Eq. \eqref{eq:jcons} together with identities and alternative forms of the equations of motion given in Appendix A, shows that the total tensor is conserved,
\begin{equation}\label{eq:conservation}
\partial_{\sigma} (T_{\rm RS}^{\sigma\tau}+T_{\rm gauge}^{\sigma\tau})=0~~~.
\end{equation}

\section{Generalized gauge invariance of the Rarita-Schwinger action}

In the most familiar gauge invariant theories, such as Abelian or non-Abelian gauge fields, the Lagrangian density
is invariant under a gauge transformation on the fields. These theories exhibit what one could term ``strong'' gauge invariance.  In a weaker form of gauge invariance, which occurs for the free  Rarita-Schwinger
equation, the Lagrangian density changes by a total derivative under a gauge transformation of the fields, and so only the action
is gauge invariant. Characteristic features of this case have been studied by  Das \cite{das1}.  We argue in this section that there is a still weaker form of gauge
invariance, obeyed by the massless Rarita-Schwinger equation with Abelian or non-Abelian gauging, in which under a gauge transformation
the Lagrangian changes by a total derivative plus terms which vanish when initial value constraints are obeyed.

In his seminal analysis of constrained systems, Dirac  \cite{dirac} classifies as ``first class'' constraints the maximal set of
constraints that have vanishing mutual Poisson brackets, and notes that ``Each of them thus leads to an arbitrary function of the time
in the general solution of the equations of motion with given initial conditions''.  Elaborating on this, he notes that
``Different solutions of the equations of motion, obtained by different choices of the arbitrary functions of the time with given
initial conditions, should be looked upon as all corresponding to the same physical state of motion, described in various way (sic) by
different choices of some mathematical variables that are not of physical significance (e.g. by different choices of the gauge
in electrodynamics or of the co-ordinate system in a relativistic theory.)''

These remarks suggest that gauge invariance, in its most general form, corresponds to an arbitrariness in the time evolution of a system,
in the sense that the future evolution of the system is not uniquely determined by the initial conditions and the Euler-Lagrange equations
following from the action principle.  Under this generalized definition, the Rarita-Schwinger equation with coupling to gauge fields
has a fermionic gauge invariance.  To see this, we note the Euler-Lagrange equations yield equations of two types.  The first are the
time evolution equations contained in Eq. \eqref{eq:eqmo}, that determine the field variables at time $t+\Delta t$ from those initially given at time $t$. Since we will see that $\psi_0$ is related by a constraint to $\vec \psi$, the evolution equations take the form
\begin{equation}\label{eq:evo1}
\vec \psi(t+\Delta t)=\vec \psi(t)+ \Delta t \frac{\partial \vec \psi}{\partial t}\Big[\vec \psi(t),\psi_0(t)\Big] + O\big((\Delta t)^2\big)~~~.
\end{equation}
The second are the primary and secondary constraints of Eqs. \eqref{eq:constraint1} and \eqref{eq:constraint2}, which constrain the
initial field values at time $t$.  If we make the gauge transformation of Eq. \eqref{eq:gaugetrans} at time $t$, with gauge parameter
$\epsilon$ of order $\Delta t$, we have seen that
the action at time $t$ changes, to first order in $\epsilon$,  according to Eq. \eqref{eq:actionchange}. So when the constraints at time $t$ are applied, the change in the action is $O\big((\Delta t)^2\Big)$.  The effect of the gauge shift $\delta_G \vec \psi(t)$ of order $\Delta t$ is
to change the time evolution equation Eq. \eqref{eq:evo1} to
\begin{align}\label{eq:evo12}
\vec \psi(t+\Delta t)=&\vec \psi(t)+\delta_G \vec \psi(t)+ \Delta t \frac{\partial \vec \psi}{\partial t}\Big[\vec \psi(t)+\delta_G \vec \psi(t),\psi_0(t)+\delta_G  \psi_0(t)\Big] + O\big((\Delta t)^2\big)\cr
=&\vec \psi(t)+\delta_G \vec \psi(t)+ \Delta t \frac{\partial \vec \psi}{\partial t}\Big[\vec \psi(t),\psi_0(t)\Big]
+O\big((\Delta t)^2\big)~~~.\cr
\end{align}
In other words, the error induced in the $\Delta t \partial \vec \psi/\partial t$ term by the gauge transformation is $O\big((\Delta t)^2\big)$.
Hence the gauge transformed variables are equally good candidates for time evolution to $t+\Delta t$,
leading to a state of the system at $t+\Delta t$ that is no longer unique; for each independent gauge change at time $t$ we get a different
set of evolved variables at time $t+\Delta t$.  One can now evolve the variables at $t+\Delta t$ to time
$t+ 2\Delta t$, subject to the constraints at time $t+\Delta t$, allowing one to make a further independent gauge transformation at time
$t+\Delta t$, and thus introducing a further non-uniqueness into the time-evolved solution.  Continuing  in this fashion, forming a Riemann
sum in the limit $\Delta t \to 0$, the $O\big((\Delta t)^2\big)$ corrections drop out, and so there is no change in the action associated with
the gauge-modified system evolution.  Thus one can
introduce arbitrary functions of time into the evolved solution, corresponding to each independent fermionic gauge transformation of
Eq. \eqref{eq:gaugetrans}. In order to get a unique time evolution path from the initial data at time $t$, one must impose a gauge
fixing condition, that selects at each time one member out of the equivalence class of equal action configurations.

Therefore, even though only the constrained action in the gauged Rarita-Schwinger theory is invariant under fermionic gauge transformations,
we shall refer to this as gauge-invariance in the sense just described, and will impose a gauge fixing condition to break this invariance.
The gauge fixing condition will also serve to give the correct helicity counting for the Rarita-Schwinger fields, as we shall see in
detail below.  We shall also discuss below how the first class constraints, which follow from repeatedly differentiating the original
primary constraints with respect to time, play a role as gauge transformation generators, in agreement with what is found in strongly
gauge invariant systems. However, we shall see that a novel feature emerges in the Rarita-Schwinger theory:  When one insists that for
each first class fermionic constraint there should be a corresponding adjoint constraint, doubling the constraint count of the theory, then the
first class constraints become second class constraints in the Dirac classification.

\section{Lagrangian analysis for left chiral spinors in two-component form }

Although we could continue with the four-component formalism to study constraints, the Hamiltonian formalism, and quantization,
it will be more convenient to first reduce the four component equation to decoupled equations for left and right chiral
components of $\psi_{\mu}^{\alpha}$ (with $\alpha$ the spinor index and with the internal symmetry index implicit).  Since these are related by symmetry, we can then focus our analysis on the two-component
equations for the left chiral component, which is the component conventionally used in formulating grand unified
models (see, e.g. \cite{adler}).

We  convert the action of Eq. \eqref{eq:action} to two-component form for the left chiral components of $\psi_{\mu}^{\alpha}$, using the Dirac matrices given in Eqs. \eqref{a2} and \eqref{a4}. Defining the two-component four vector spinor $\Psi_{\mu}^{\alpha}$ and its adjoint $\Psi_{\mu \alpha}^{\dagger}$ by
\begin{align}\label{eq:Psidef}
 P_L \psi_\mu^{\alpha}=&\left( \begin{array} {c}
 \Psi_\mu^{\alpha}  \\
 0 \\  \end{array}\right)~,~~ \mu=0,1,2,3  ~,~~ \alpha=1,2~~~,\cr
 \psi_{\mu\alpha}^{\dagger} P_L=&\Big(\Psi_{\mu \alpha}^{\dagger}~~~0\Big)~~~,\cr
\end{align}
the  action decomposes into uncoupled left and right chiral parts. The left chiral part, with spinor indices $\alpha$ suppressed,  is given by
\begin{equation}\label{eq:leftaction}
S(\Psi_{\mu})=\frac{1}{2}\int d^4x  [-\Psi_{0}^{\dagger} \vec \sigma \cdot \vec D \times \vec {\Psi}
+\vec {\Psi}^{\dagger} \cdot \vec \sigma \times \vec D \Psi_{0}
+\vec{\Psi} ^{\dagger} \cdot \vec D \times \vec \Psi - \vec{\Psi}^{\dagger} \cdot \vec \sigma \times D_{0} \vec{\Psi}]~~~.
\end{equation}
Varying with respect to $\vec{\Psi}^{\dagger}$ we get the Euler-Lagrange equation
\begin{equation}\label{eq:euler}
0=\vec V \equiv \vec \sigma \times \vec D \Psi_{0} + \vec D \times \vec{\Psi}-\vec \sigma \times D_{0}\vec {\Psi}~~~,
\end{equation}
while varying with respect to $\Psi_{0}^{\dagger}$ we get the primary constraint \big(given in four-component form in Eq. \eqref{eq:constraint1}\big)
\begin{equation}\label{eq:chi}
0=\chi \equiv \vec \sigma \cdot \vec D \times \vec{\Psi}~~~.
\end{equation}
A second primary constraint follows from the fact that the action has no dependence on $d\Psi_0^{\dagger}/dt$, which implies that
the  momentum conjugate to $\Psi_0^{\dagger}$ vanishes identically,
\begin{equation}\label{eq:p0}
P_{\Psi_0^{\dagger}}=0~~~.
\end{equation}

Contracting $\vec V$ with $\vec \sigma$ and with $g^{-1}\vec D$, and using the covariant derivative relations of Eq. \eqref{a14}, we
get respectively
\begin{align}\label{eq:omegatheta1}
\vec \sigma \cdot \vec V= &2i \theta +\chi~~~,\cr
g^{-1}\vec D \cdot \vec V=&i \omega+g^{-1} D_0 \chi~~~,\cr
\end{align}
with
\begin{align}\label{eq:omegatheta2}
\theta\equiv&\vec \sigma \cdot \vec D \Psi_0 - D_0 \vec \sigma \cdot \vec \Psi~~~,\cr
\omega\equiv& \vec \sigma \cdot \vec B \Psi_0 -(\vec B + \vec \sigma \times \vec E) \cdot \vec \Psi~~~.\cr
\end{align}
Since the Euler-Lagrange equations imply that $\vec V$ and $\chi$ vanish for all times, we learn that $\theta$ and $\omega$
vanish also for all times.  Since $\theta$ involves a time derivative, its vanishing is just one component of the equation
of motion for $\Psi_{\mu}$.  But $\omega$ involves no time derivatives, so it is a secondary constraint that relates $\Psi_0$  to
$\vec \Psi$ \big(given in four-component form in Eq. \eqref{eq:constraint2}\big).   For each of the above equations, there is a corresponding relation for the adjoint quantity.

The equation of motion $\vec V=0$ can be written in a simpler form by using the identities of Eqs. \eqref{a10} and \eqref{a11} as
follows.  Using Eq. \eqref{a10} to simplify $0=\vec \sigma \times \vec V- i \vec V$, we get an equation for $D_0 \vec \Psi$,
\begin{equation}\label{eq:d0psi1}
D_0\vec \Psi= \vec D \Psi_0+ \frac{1}{2} [-\vec \sigma \times (\vec D \times \vec{\Psi}) + i \vec D \times \vec \Psi]~~~.
\end{equation}
A further simplification can be achieved by incorporating the primary constraint $\chi=0$, through applying Eq. \eqref{a11} to
$\vec A=\vec D \times \vec \Psi$,
\begin{equation}\label{eq:simpli}
0=\vec \sigma ~\chi = \vec \sigma ~\vec \sigma \cdot(\vec D \times \vec \Psi)=\vec D \times \vec \Psi - i \vec \sigma \times (\vec D \times \vec \Psi)~~~.
\end{equation}
Using this to replace the first term in square brackets in Eq. \eqref{eq:d0psi1} we get the alternative
form of the equation of motion, valid when the constraint $\chi=0$ is satisfied,
\begin{equation}\label{eq:d0psi2}
D_0 \vec \Psi=\vec D \Psi_0+ i \vec D \times \vec \Psi~~~.
\end{equation}

Writing the gauge field interaction terms in Eq. \eqref{eq:leftaction} in the form
\begin{equation}\label{eq:interaction}
S_{\rm int} (\Psi_{\mu})= \frac{g}{2} \int d^4 x (A_0^B J^{B0}+\vec A^B \cdot {\vec J}^B)~~~,
\end{equation}
we find the left chiral contribution to the current of Eq. \eqref{eq:gaugemo} in the form
\begin{align}\label{eq:currents}
J^{A0}=&-\vec{\Psi}^{\dagger} t_A \cdot \vec \sigma \times \vec{\Psi}~~~,\cr
\vec{J}^A=& \Psi_0^{\dagger} t_A \vec \sigma \times \vec \Psi+ \vec {\Psi}^{\dagger} \times \vec \sigma t_A \Psi_0 -
\vec{\Psi}^{\dagger} \times t_A \vec{\Psi}~~~.\cr
\end{align}
Replacing $t_A$ by $-i$, we find the corresponding singlet current in the form
\begin{align}\label{eq:currents1}
J^{0}=&i\vec{\Psi}^{\dagger}  \cdot \vec \sigma \times \vec{\Psi}~~~,\cr
\vec{J}=&-i (\Psi_0^{\dagger}  \vec \sigma \times \vec \Psi+ \vec {\Psi}^{\dagger} \times \vec \sigma \Psi_0 -
\vec{\Psi}^{\dagger} \times \vec{\Psi})~~~.\cr
\end{align}
For the energy integral computed from the left chiral part of the the stress-energy tensor of Eq. \eqref{eq:rstensor}, we find
\begin{equation}\label{eq:graven}
H=-\int d^3x T_{RS}^{00}= -\frac{1}{2} \int d^3x \vec{\Psi}^{\dagger} \cdot \vec{D} \times \vec{\Psi}~~~.
\end{equation}

To conclude this section, we verify that the action of Eq. \eqref{eq:leftaction} has a fermionic gauge invariance on the constraint surface
$\omega=0~,~~ \omega^{\dagger}=0$, as already seen in covariant form following Eq. \eqref{eq:gaugetrans}.   Letting $\epsilon$ be a general space and time dependent two-component spinor, we introduce the fermionic gauge changes
\begin{align}\label{eq:gaugedef}
\vec{\Psi} \to & \vec {\Psi} + \delta_G \vec \Psi ~~,~~\delta_G \vec \Psi\equiv \vec D \epsilon~~~,\cr
\Psi_0 \to & \Psi_0 + \delta_G \Psi_0~~,~~  \delta_G \Psi_0 \equiv  D_0 \epsilon~~~, \cr
\end{align}
and their adjoints,
which are the left chiral form of the gauge change of Eq. \eqref{eq:gaugetrans}.   Substituting this into Eq. \eqref{eq:leftaction}, integrating by parts where needed,
and using Eqs. \eqref{a14} to simplify commutators of
covariant derivatives, we find that Eq. \eqref{eq:actionchange} takes the two-component spinor form
\begin{equation}\label{eq:actionchange1}
\delta_G S(\Psi_{\mu}) = \frac{1}{2}ig \int d^4 x (\omega^{\dagger}\epsilon -\epsilon^{\dagger} \omega)~~~.
\end{equation}
Hence the  action on the constraint surface $\omega=\omega^{\dagger}=0$ has a fermionic gauge invariance.
Another gauge invariant, on the constraint surface $\chi=\chi^{\dagger}=0$,  is the fermion number, given
by the space integral of the time component of the singlet current of Eq. \eqref{eq:currents1}, $\int d^3x J^0 $,
which has the gauge variation
\begin{equation}\label{eq:fermionnumbervar}
\delta_G\int d^3x J^0=-i\int d^3x (\epsilon^{\dagger}\chi+\chi^{\dagger}\epsilon)~~~.
\end{equation}
However, neither the equation of motion, the constraints $\chi$ and $\omega$, the non-Abelian ``charge''  $\int d^3x J^{B0}$,  nor the integrated Hamiltonian $H$ are gauge invariant
in the interacting case. Using $\delta_G$ to denote gauge variations, we have
\begin{align}\label{eq:nongaugeinv}
\delta_G \vec V=&-ig(\vec B + \vec \sigma \times \vec E) \epsilon~~~,\cr
\delta_G \theta =& -ig \vec \sigma \cdot \vec E \epsilon~~~,\cr
\delta_G \chi=&-ig \vec \sigma \cdot \vec B \epsilon~~~,\cr
\delta_G \omega=& \vec \sigma \cdot \vec B D_0 \epsilon - (\vec B+ \vec \sigma \times \vec E) \cdot \vec D \epsilon~~~,\cr
\delta_G \int d^3x J^{B0}=&g\int d^3x \Big( \epsilon^{\dagger} [\vec A, t_B] \cdot \vec \sigma \times \vec \Psi
+ \vec \Psi^{\dagger} \times \vec \sigma \cdot [t_B,\vec A] \epsilon  \Big)~~~,\cr
\delta_G H=& \frac{1}{2}ig \int d^3x (\vec \Psi^{\dagger} \cdot \vec B \epsilon - \epsilon^{\dagger} \vec B \cdot \vec \Psi) ~~~.\cr
\end{align}
The last line shows that in the interacting case,  there is no total energy integral that is invariant
under the fermionic gauge transformation, much as there is no generally covariant gravitational total energy.
The {\it only} global fermionic gauge invariants are the action integral, and the fermion number integral.

To break the gauge invariance we must introduce an additional constraint, in the form
\begin{equation}\label{eq:constraint}
f(\vec \Psi)=0~~~,
\end{equation}
with $f$ a scalar function of its argument,  such as $f=\vec{D} \cdot \vec \Psi$ (a gauge covariant radiation gauge analog) or
$f=(\vec B +\vec \sigma \times \vec E)\cdot \vec \Psi$ \big(which by Eq. \eqref{eq:omegatheta2} corresponds, when $\vec \sigma \cdot \vec B$ is invertible, to $\Psi_0=0$.\big). This constraint, together with the $\chi$ constraint, leaves one independent
two-component spinor of the original three in $\vec \Psi$, corresponding to the physical massless Rarita-Schwinger
modes propagating in the gauge field background.  We will limit ourselves to considering linear constraints of the general form
\begin{equation}\label{eq:constraintform}
f=\vec L \cdot \vec \Psi~~~,
\end{equation}
so that for gauge covariant radiation gauge we have $\vec L = \vec D$, and for $\Psi_0=0$ gauge we have $\vec L=\vec B +\vec \sigma \times \vec E$.
Both of these choices of $\vec L$  play a special role in our analysis, so we shall examine them in more detail.

We consider first the gauge covariant radiation gauge.
We note that since
\begin{equation}\label{eq:chisimp}
\vec \sigma \cdot \vec D \vec \sigma \cdot \vec{\Psi}=\vec D \cdot \vec{\Psi}+ i\chi,
\end{equation}
the primary constraint $\chi=0$ implies that
\begin{equation}\label{eq:chisimp1}
\vec \sigma \cdot \vec D \vec \sigma \cdot \vec{\Psi}=\vec D \cdot \vec{\Psi}~~~.
\end{equation}
Hence when $\vec \sigma \cdot \vec D$ is invertible, which is expected in a perturbation expansion in the
gauge coupling $g$, the covariant radiation gauge constraint $\vec D \cdot \vec \Psi=0$ implies that
\begin{equation}\label{eq:sigmaconstraint}
\vec \sigma \cdot \vec \Psi=0~~~.
\end{equation}
Conversely, Eqs. \eqref{eq:chisimp} and \eqref{eq:chisimp1} show that $\vec D \cdot \vec \Psi=0$  and $\vec \sigma \cdot \vec \Psi=0$
together imply the primary constraint $\chi=0$, and also $\vec \sigma \cdot \vec \Psi=0$ and $\chi=0$ imply $\vec D \cdot \vec \Psi=0$.

We next note that on a given initial time slice, covariant radiation gauge is attainable.  Under the gauge
transformation of Eq. \eqref{eq:gaugedef}, we see that
\begin{equation}\label{eq:radgaugechange}
\vec D \cdot \vec \Psi \to \vec D \cdot \vec \Psi+ (\vec D)^2  \epsilon~~~.
\end{equation}
Hence when $(\vec D)^2$ is invertible, which we expect to be true in a perturbative
sense, then we can invert $(\vec D)^2 \epsilon = - \vec D \cdot \vec \Psi$, to find
a gauge function $\epsilon$ that brings a general $\vec \Psi$ to covariant radiation
gauge.  Since
\begin{equation}\label{eq:sigmadsquared}
(\vec \sigma \cdot \vec D)^2=(\vec D)^2+g \vec \sigma \cdot \vec B~~~,
\end{equation}
the conditions for $\vec \sigma \cdot \vec D$ to be invertible, and
for $(\vec D)^2$ to be invertible, are  related.  For generic non-Abelian
gauge fields both of these operators should be invertible, but there will be
isolated gauge field configurations for which $\vec \sigma \cdot \vec D$ has zeros.

However, although covariant radiation gauge can be imposed on any time slice, it is
not preserved by the equation of motion for $\vec \Psi$.  To see this, let us consider
the simplified case in which the gauge potential is specialized to $A_0=0$ and $\partial_0 \vec A=0$,
so that only a static $\vec B$ field is present.  From Eq. \eqref{eq:d0psi2} we have
\begin{equation}\label{eq:d0radgauge}
\partial_0(\vec D \cdot  \vec \Psi)=(\vec D)^2 \Psi_0+ g \vec B \cdot \vec \Psi = \big((\vec D)^2 (\vec \sigma \cdot \vec B)^{-1}+g\big) \vec B \cdot \vec \Psi\neq 0~~~.
\end{equation}
Hence at each infinitesimal time step, we must make a further infinitesimal gauge transformation
to maintain the gauge covariant radiation gauge condition.  This means that we {\it cannot} use
the $\theta=0$ time development equation
\begin{equation}\label{eq:psi0constraint}
\vec \sigma \cdot \vec D \Psi_0 = D_0 \vec \sigma \cdot \vec \Psi~~~,
\end{equation}
together with  Eq. \eqref{eq:sigmaconstraint}, and the assumption that $\vec \sigma \cdot \vec D$ is invertible,
to conclude that covariant radiation gauge also implies that $\Psi_0=0$.

We can, however, impose $\Psi_0=0$ as an alternative gauge condition.
Under the gauge  transformation of Eq. \eqref{eq:gaugedef},
$\Psi_0$ can be reduced to zero by taking
\begin{align}\label{eq:psi0trans1}
\epsilon=& -V(t,\vec x)^{-1} \int_0^t du V(u,\vec x) \Psi_0(u,\vec x)~~~,\cr
V(t,\vec x)=&\exp\left(g\int_0^t du A_0(u,\vec x)\right)~~~.\cr
\end{align}

Once $\Psi_0$ has been gauged to zero, we can conclude from Eq. \eqref{eq:omegatheta2}
that $(\vec B +\vec \sigma \times \vec E)\cdot \vec \Psi=0$.  Just as we found for
covariant radiation gauge,  $\Psi_0=0$ gauge
is not preserved by the equation of motion for $\vec \Psi$.  To see this, let us again consider
the simplified case in which the gauge potential is specialized to $A_0=0$ and $\partial_0 \vec A=0$,
so that only a static $\vec B$ field is present.  From Eq. \eqref{eq:d0psi2} we have
\begin{equation}\label{eq:psi0trans2}
\partial_0 (\vec B \cdot \vec \Psi)= i\vec B \cdot \vec D \times \vec \Psi \neq 0~~~.
\end{equation}
So again, at each infinitesimal time step, we must make a further infinitesimal gauge transformation
to maintain the gauge condition $\Psi_0=0$.

To summarize this discussion of gauge fixing, we see that when the gauge fields are nonzero, we can
have either $\vec D\cdot \vec \Psi=0$ or $\Psi_0=0$, but not both.  In the free field case, these
two conditions can be imposed simultaneously and are preserved by the time evolution
of $\vec \Psi$, so we recover the constraints  $\vec \nabla \cdot \vec{\psi}=0$, $\vec \sigma \cdot \vec{\psi}=0$, and
$\psi_0=0$ used in the discussion of \cite{dasb} for the free Rarita-Schwinger case,

\section{Propagation of a Rarita-Schwinger field in an external Abelian gauge field:  absence of superluminal propagation}

We  specialize now to the case of a Rarita-Schwinger spinor propagating in an external Abelian gauge field, as studied by
Velo and Zwanziger \cite{velo}.  For an Abelian gauge field,
\begin{equation}\label{eq:sigbinversion}
\frac{1}{\vec \sigma \cdot \vec B}=\frac{\vec \sigma \cdot \vec B}{(\vec B)^2}~~~,
\end{equation}
and so $\vec \sigma \cdot \vec B$ is invertible as long as $(\vec B)^2>0$, which we assume.   Provided the Lorentz invariant expression $(\vec B)^2-(\vec E)^2$ is positive, $(\vec B)^2$ will be positive in any Lorentz frame.   In discussing undamped wave propagation we will not use the inequality $(\vec B)^2-(\vec E)>0$, but in treating damped longitudinal mode propagation in Appendix B, we
will assume that $(\vec E)^2/(\vec B)^2$ is small, as motivated by the fact that when $(\vec E)^2$ is of order
 $(\vec B)^2$ the vacuum is highly unstable against pair creation. \big(Strictly speaking, the vacuum is stable against
pair production only when $\vec E \cdot \vec B=0$ and $(\vec B)^2 -(\vec E)^2 >0$, that is, when there is
a Lorentz frame in which the Abelian field has vanishing $\vec E$ \cite{schwinger}.\big)

Given that $(\vec B)^2>0$, we can solve the constraint $\omega=0$ of Eq. \eqref{eq:omegatheta2} for
$\Psi_0$, giving
\begin{equation}\label{eq:solve1}
\Psi_0=\frac{\vec Q  \cdot \vec \Psi}{(\vec B)^2}~~~,
\end{equation}
where we have defined
\begin{equation}\label{eq:qdef}
\vec Q\equiv  \vec \sigma \cdot \vec B (\vec B + \vec \sigma \times \vec E)=
\vec B\times \vec E + \vec B \vec \sigma \cdot (\vec B +i \vec E)-i \vec B \cdot \vec E \vec \sigma~~~.
\end{equation}
Substituting the solution for $\Psi_0$ into Eq. \eqref{eq:d0psi2}, we get an equation of motion for $\vec \Psi$
by itself,
\begin{equation}\label{eq:d0psi3}
D_0 \vec \Psi=\vec D \frac{\vec Q  \cdot \vec \Psi}{(\vec B)^2}+ i \vec D \times \vec \Psi~~~.
\end{equation}

To determine the wave propagation velocity in the neighborhood of a point $\vec x_*$, we need to calculate the equation for
the wavefronts, or characteristics, at that point.   Writing the first order  Eq. \eqref{eq:d0psi3} in the form
\begin{equation}\label{eq:d0psi4}
\partial_0 \vec \Psi=\vec \nabla \frac{\vec Q_*  \cdot \vec \Psi}{(\vec B_*)^2}+ i \vec  \nabla\times \vec \Psi+
\vec \Delta[\vec \Psi, \vec x_*,\vec x]~~~,
\end{equation}
with $\vec B_*$ and $\vec Q_*$ the values of the respective quantities at $\vec x_*$,
we see that $\vec \Delta[\vec \Psi, \vec x_*,\vec x]$ involves no first derivatives of $\vec \Psi$ at  $\vec x_*$, and so is not needed \cite{courant},
 \cite{madore} for determining the wavefronts of  Eq. \eqref{eq:d0psi2}.  The reason  is that when taking an infinitesimal line integral
of Eq. \eqref{eq:d0psi4}, according to
\begin{equation}\label{eq:lineint}
\lim_{\delta \to 0} \int_{-\delta}^{\delta} [\partial_0 \vec \Psi = ...]~~~,
\end{equation}
discontinuities across wavefronts contribute through the first derivative terms, but when the external fields are smooth
the term $\vec \Delta[\vec \Psi, \vec x_*,\vec x]$  makes a vanishing contribution as $\delta \to 0$.
Dropping $\vec \Delta$, and multiplying through by $(\vec B_*)^2$, we get the equation determining the wavefronts in the form
\begin{equation}\label{eq:d0psifinal}
(\vec B_*)^2 \partial_0 \vec \Psi=\vec \nabla \vec Q_*  \cdot \vec \Psi+ i (\vec B_*)^2 \vec  \nabla\times \vec \Psi~~~.
\end{equation}
By similar reasoning, the constraint $\chi$ can be simplified, for purposes of determining the wavefronts, by replacing
$\vec D$ by $\vec \nabla$, giving
\begin{equation}\label{eq:newchi}
0=\vec \sigma \cdot \vec \nabla   \times \vec{\Psi}~~~,
\end{equation}
and the covariant radiation gauge condition similarly simplifies to
\begin{equation}\label{eq:newradgauge}
0=\vec \nabla \cdot \vec{\Psi}~~~.
\end{equation}

Since these are now linear equations with constant coefficients, the solutions are plane waves, and without loss of generality we can take the negative $z=x_3$ axis as the direction of wave propagation.  So making the Ansatz
\begin{equation}\label{eq:ansatz}
\vec \Psi= \vec C \exp(i \Omega t + i K z)~~~,
\end{equation}
 Eq. \eqref{eq:d0psifinal} for the wavefronts or characteristics takes the form
\begin{equation}\label{eq:ceq}
0=\vec F\equiv (\vec B_*)^2 \Omega \vec C- K \hat z\vec Q_*  \cdot \vec C- i (\vec B_*)^2  K \hat z \times \vec C~~~,
\end{equation}
with $\hat z$ a unit vector along the $z$ axis. Similarly, the constraint Eq. \eqref{eq:newchi} becomes an admissability
condition on $\vec C$,
\begin{equation}\label{eq:newchi1}
0=\vec \sigma \cdot \hat z   \times \vec C~~~,
\end{equation}
and the gauge fixing condition $\vec \nabla \cdot \vec \Psi=0$ imposes the further condition on $\vec C$
\begin{equation}\label{eq:newgaugefix}
0=\hat z \cdot \vec C~~~,
\end{equation}
but for the moment we will analyze the solutions of Eq. \eqref{eq:ceq} without assuming
this additional constraint.

Writing $F_m$ as a matrix times $C_n$ (and dropping the subscripts $*$, which are implicit from here on) we have
\begin{align}\label{eq:matrix}
F_m=&N_{mn}C_n~~~,\cr
N_{mn}=&(\vec B)^2 \Omega \delta_{mn}- K \delta_{m3} Q_n  - i (\vec B)^2  K \epsilon_{m3n}~~~.\cr
\end{align}
The equation for the characteristics is now
\begin{equation}\label{eq:char}
{\rm det} (N)=0~~~,
\end{equation}
since this is the condition for Eq. \eqref{eq:ceq} to have a solution with nonzero $\vec C$.  However, since evaluation of
the determinant shows that it factorizes into blocks that determine $C_{1,2}$ and a  block that determines $C_3$, a simpler
way to proceed is to work directly from the equations $F_m=0$, which decouple in a corresponding way. Calculating from
Eq. \eqref{eq:ceq},  we find
\begin{align}\label{eq:fcomp}
0=&F_1^{\uparrow,\downarrow}=(\vec B)^2 \big(\Omega C_1^{\uparrow,\,\downarrow}+iK C_2^{\uparrow,\,\downarrow}\big)~~~,\cr
0=&F_2^{\uparrow,\,\downarrow}=(\vec B)^2 \big(\Omega C_2^{\uparrow,\,\downarrow}-iK C_1^{\uparrow,\,\downarrow}\big)~~~,\cr
0=&F_3^{\uparrow,\,\downarrow}=(\vec B)^2 \Omega C_3^{\uparrow,\,\downarrow}- K (\vec Q \cdot \vec C)^{\uparrow,\,\downarrow}~~~,\cr
\end{align}
where $\uparrow,\, \downarrow$ indicate the up and down spinor components, labeled in Eq. \eqref{eq:Psidef} by $\alpha=1,\,2$. Similarly,
the constraint Eq. \eqref{eq:newchi1} becomes $0= -\sigma_1 C_2 +\sigma_2 C_1$, that is
\begin{align}\label{eq:newchi2}
C_2^{\uparrow}=&iC_1^{\uparrow}~~~,\cr
C_2^{\downarrow}=&-i C_1^{\downarrow}~~~,\cr
\end{align}
with no corresponding condition on $C_3^{\uparrow,\,\downarrow}$.

The first two lines of Eq. \eqref{eq:fcomp} together with Eq. \eqref{eq:newchi2} have the solution
\begin{align}\label{eq:c12soln}
C_1^{\uparrow}=&C~,~~C_2^{\uparrow}=iC~,~~\Omega=K~~~,\cr
C_1^{\downarrow}=&C~,~~C_2^{\downarrow}=-iC~,~~\Omega=-K~~~,\cr
\end{align}
with C arbitrary, corresponding to waves with velocity of magnitude $|\Omega/K|=1$.
The effect of a gauge change $\vec \Psi \to \vec \Psi + \vec D \epsilon$, $\epsilon= \big(E/(iK)\big) \exp(i \Omega t + i K z)$ on
the characteristics is to shift $C_3^{\uparrow,\downarrow} \to C_3^{\uparrow,\downarrow} + E^{\uparrow,\downarrow} $, thus replacing the third line of Eq. \eqref{eq:fcomp} by the two-component
spinor equation $\big((\vec B)^2 \Omega-K Q_3\big) (C_3+E)- K (Q_1 C_1+ Q_2 C_2)=0$.  If we
choose $E$ to cancel the term $K (Q_1 C_1+ Q_2 C_2)$, that is,
\begin{equation}\label{eq:ecancel}
E=\frac{K(Q_1C_1+Q_2C_2)}{(\vec B)^2\Omega-KQ_3}~~~,
\end{equation}
then this equation is solved by $C_3=0$ (corresponding
to enforcing the covariant radiation gauge conditions $\vec \nabla \cdot \vec \Psi = \vec \sigma \cdot \vec \Psi=0$),
and the demonstration that there is no superluminal propagation is complete.
In Appendix B, we continue this discussion without assuming a specific gauge condition, and show that there are still no propagating
modes with superluminal velocities.  That is, the purely longitudinal mode with $C_1=C_2=0, C_3 \neq 0$, which is eliminated by the
gauge fixing condition, does not propagate superluminally.

\section{Canonical momenta, classical brackets, and gauge generators}

We return now to the general non-Abelian gauge field case, and introduce the canonical momentum conjugate to $\vec \Psi$,
defined by
\begin{equation}\label{eq:canmom}
\vec P = \frac{\partial^L S}{\partial (\partial_0 \vec{\Psi}) } = \frac{1}{2} \vec{\Psi}^{\dagger} \times \vec \sigma~~~,
\end{equation}
which can be solved for $\vec{\Psi}^{\dagger}$ using the final line of Eq. \eqref{a11},
\begin{equation}\label{eq:psidagger}
\vec{\Psi}^{\dagger}=i \vec P - \vec P \times \vec \sigma~~~.
\end{equation}
We will use Eq. \eqref{eq:psidagger} when computing classical brackets involving $\vec{\Psi}^{\dagger}$ using the formula of
Eq. \eqref{a17}.
Eq. \eqref{eq:canmom} can be written as an explicit matrix relation for the  six components of $\vec P$ and $\vec{\Psi}^{\dagger}$,
\begin{equation}\label{eq:matrixform}
\left( \begin{array} {c} P_1^{\uparrow}\\P_1^{\downarrow}\\
P_2^{\uparrow} \\P_2^{\downarrow}\\P_3^{\uparrow}\\ P_3^{\downarrow}\\
\end{array}\right)=\frac{1}{2}\left( \begin{array} {cccccc}
 0  & 0 & 1 &0   &0   & -i   \\
 0  & 0 &0  & -1  & i  & 0   \\
-1   &0  &0  &0   &0   &1    \\
 0  &1  &0  &0   &1   & 0   \\
 0  &i  &0  & -1  &0   & 0   \\
 -i  &0 &-1  &0   & 0  & 0   \\
\end{array} \right)
\left( \begin{array} {c} \Psi_1^{\dagger\uparrow}\\ \Psi_1^{\dagger\downarrow}\\
\Psi_2^{\dagger\uparrow} \\ \Psi_2^{\dagger\downarrow}\\ \Psi_3^{\dagger\uparrow}\\ \Psi_3^{\dagger\downarrow}\\
\end{array}\right)~~~,
\end{equation}
showing that they are related by an anti-self-adjoint matrix with determinant $-1/16$.

The four constraints introduced in Sec. 4 are
\begin{align}\label{eq:phicon}
\phi_1=&P_{\Psi_0^{\dagger}}~~~,\cr
\phi_2=&(\vec \sigma \cdot \vec B)^{-1} \omega= \Psi_0-(\vec \sigma \cdot \vec B)^{-1} (\vec B+\vec \sigma \times \vec E)
\cdot \vec\Psi~~~,\cr
\phi_3=&\chi=\vec \sigma \cdot \vec D \times \vec \Psi~~~,\cr
\phi_4=&\vec L \cdot \vec \Psi~~~.\cr
\end{align}
In writing these we are assuming that $\vec \sigma \cdot \vec B$ is invertible in the non-Abelian case.  We are writing the gauge fixing condition as  a general  linear gauge fixing constraint $\vec L\cdot \vec \Psi$, so
as to keep track of which terms in the final answers arise from gauge fixing, which is not evident if we specialize by replacing $\vec L$ by
$\vec D$ or $\vec B + \vec \sigma \times \vec E$.  The constraints of Eq. \eqref{eq:phicon}, including the gauge fixing
constraint $\phi_4$, are all first class in the Dirac classification, since they have vanishing mutual classical brackets.  This is
a consequence of the fact that starting with a constraint depending on $\vec \Psi$ but not on $\vec \Psi^{\dagger}$, and taking an arbitrary
number of time derivatives, one still has a constraint depending only on $\vec \Psi$.

To preserve the adjoint properties of the Rarita-Schwinger equation, for each of these four constraints we must impose
 a corresponding adjoint constraint.  Using Eq. \eqref{eq:psidagger} to express $\vec{\Psi}^{\dagger}$
in terms of $\vec P$, we write these as
\begin{align}\label{eq:chidefs}
\chi_1=&(P_{\Psi_0^{\dagger}})^{\dagger}=-P_{\Psi_0}~~~,\cr
\chi_2=&\omega^{\dagger}(\vec \sigma \cdot \vec B)^{-1}=\Psi_0^{\dagger}- \vec P \cdot [i(\vec B+\vec \sigma \times \vec E)
-  \vec \sigma \times (\vec B+\vec \sigma \times \vec E)](\vec \sigma \cdot \vec B)^{-1}~~~,\cr
\chi_3=&\chi^{\dagger}=2 \vec P \cdot \overleftarrow D ~~~,\cr
\chi_4=& \vec{\Psi}^{\dagger} \cdot \overleftarrow{L}= \vec P \cdot (i \overleftarrow{L} -  \vec \sigma \times \overleftarrow{L})~~~.\cr
\end{align}
(The reason for the minus sign in the definition  $P_{\Psi_0^{\dagger}}=-P_{\Psi_0}^{\dagger}$ will be given in Sec. 8 where we discuss the Hamiltonian
form of the equations.)  The constraints $\phi_a$ are implicitly $2n$ component column vectors, and the adjoint constraints $\chi_a$ are implicitly $2n$ component row vectors,
with $2n$ arising from the product of a factor of 2 for the two implicit spinor indices, and a factor of $n$ for the $n$  implicit $SU(n)$  internal symmetry indices.

When $\vec L=\vec D$, we see that $\phi_4$ becomes $\phi_4=\vec D \cdot \vec \Psi$, and
$\chi_4$ becomes $\chi_4=i\vec P \cdot \overleftarrow{D} - \vec P \cdot \vec \sigma \times \overleftarrow{D} = (i/2) \chi_3 -\vec P \cdot \vec \sigma \times \overleftarrow{D}$.
So a special feature of covariant radiation gauge, which will be exploited later, is that the constraints $\phi_3,\,\phi_4$ are contractions
of $\vec \sigma \times \vec D$ and $\vec D$ with $\vec \Psi$, and the constraints $\chi_3,\,\chi_4$ are contractions of linear combinations
of the duals $\overleftarrow{D}$ and $\vec \sigma \times \overleftarrow{D}$ with $\vec P$.  That is, in covariant radiation gauge the
constraint spaces selected by $\chi_3,\chi_4$ and $\phi_3,\phi_4$ are similar.

We can now compute the classical brackets of the constraints.  We see that the brackets of the $\phi$s and $\chi$s vanish among themselves,
\begin{align}\label{eq:vanishbracks}
[\phi_a,\phi_b]_C=&0~~~,\cr
[\chi_a,\chi_b]_C=&0~~~,\cr
a,b=&1,...,4~~~.\cr
\end{align}
On the other hand, the brackets of the $\phi$s with the $\chi$s give a nontrivial matrix of brackets $M$, which has a nonvanishing determinant,
\begin{align}\label{eq:nonvanishbracks}
M_{ab}(\vec x, \vec y) \equiv &  [\phi_a(\vec x),\chi_b(\vec y)]_C \neq 0~~~,\cr
\det{M} \neq & 0~~~.\cr
\end{align}
Thus, in terms of the Dirac classification, the original first class constraints $\phi_a$ have become second class, not from
adding new constraints that follow from differentiation with respect to time or from imposing gauge fixing conditions, but rather
from adjoining the adjoint set of constraints.  This is a feature of the Rarita-Schwinger constrained fermion system that
has no analog in the familiar constrained boson systems such as gauge fields.

Evaluating the brackets shows that $M$ has the general form
\begin{equation}\label{eq:pstructure}
M=\left( \begin{array} {cccc}
 0&-1&0&0 \\
 1&{\cal U}&{\cal S}&{\cal T} \\
 0&{\cal V}&{\cal A}&{\cal B} \\
 0&{\cal W}&{\cal C}&{\cal D} \\
 \end{array} \right)~~~,
\end{equation}
where in the $SU(n)$ gauge field case, each entry in $M$ is a $2n\times 2n $ matrix (corresponding to the fact that $\phi_a$ is implicitly a $2n$ component
column vector, and $\chi_b$ is implicitly a $2n$ component row vector).  Evaluating
$\det{M}$ by a cofactor expansion with respect to the elements of the two unit matrices $\pm 1$, we see that
the submatrices ${\cal U},\, {\cal S},\, {\cal T},\, {\cal V},\,{\cal W}$ do not contribute, and we have
\begin{align}\label{eq:detp}
\det{M}=&\det{N}~~~\cr
N= & \left(\begin{array} {cc}
  {\cal A}&{\cal B} \\
 {\cal C}&{\cal D} \\
 \end{array} \right)   ~~~.\cr
 \end{align}
So we need to only evaluate the brackets $M_{33}={\cal A}$, \, $M_{34}={\cal B}$,\,
$M_{43}={\cal C}$,\, $M_{44}={\cal D}$, giving
\begin{align}\label{eq:abcdformulas}
{\cal A}=& -2ig \vec \sigma \cdot \vec B(\vec x) \delta^3(\vec x-\vec y)~~~, \cr
{\cal B}=&-2 \vec D_{\vec x}\cdot \vec L_{\vec x} \delta^3(\vec x-\vec y)~~~, \cr
{\cal C}=& 2 \vec{L}_{\vec x}\cdot \vec D_{\vec x} \delta^3(\vec x-\vec y)~~~, \cr
{\cal D}=& \big(i (\vec{L}_{\vec x})^2+\vec \sigma \cdot(\vec L_{\vec x} \times \vec L_{\vec x})\big) \delta^3(\vec x-\vec y)~~~. \cr
\end{align}
When $\vec L=\vec D$, these become
\begin{align}\label{eq:abcdformulas1}
{\cal A}=& -2ig \vec \sigma \cdot \vec B(\vec x) \delta^3(\vec x-\vec y)~~~, \cr
{\cal B}=&-2 (\vec D_{\vec x})^2 \delta^3(\vec x-\vec y)~~~, \cr
{\cal C}=& 2 (\vec D_{\vec x})^2 \delta^3(\vec x-\vec y)~~~, \cr
{\cal D}=& i\big( (\vec{D}_{\vec x})^2-g \vec \sigma \cdot \vec B({\vec x})\big) \delta^3(\vec x-\vec y)~~~. \cr
\end{align}

Reflecting the fact that the $\phi_a$ and $\chi_a$ are adjoints of one another, together with
the fact that the matrix relating  $\vec \Psi^{\dagger} $ to $\vec P$ is anti-self-adjoint \big(see Eq. \eqref{eq:matrixform}\big),
these matrix elements obey the adjoint relations
\begin{equation}\label{adjointrels}
M_{ab}(\vec x, \vec y)^{\dagger}=-M_{ba}(\vec y, \vec x)~~~~.
\end{equation}
Applications of these bracket and determinant calculations will be made in the
next two sections.

To conclude this section, we note that the constraints $\chi,\, \chi^{\dagger},\, P_{\Psi_0},\,
P_{\Psi_0^{\dagger}}$ play the role of gauge transformation generators. For example,
we have (with common time argument $t$ suppressed)
\begin{align}\label{eq:gaugegenex}
\left[\int d^3x \frac{1}{2}\chi^{\dagger}(\vec x)  \epsilon(\vec x), \vec \Psi(\vec y)\right]_C=&\vec  D_{\vec y}\, \epsilon(\vec y)~~~,\cr
\left[-\int d^3x P_{\Psi_0}(\vec x) D_0 \epsilon(\vec x), \Psi_0(\vec y)\right]_C=& D_{0\vec y}\, \epsilon(\vec y)~~~.\cr
\end{align}
So the fermionic gauge transformation is a canonical transformation.

\section{Path integral quantization in $\Psi_0=0$  gauge}

In studying quantization, we will specialize to the case where the external gauge potentials, and hence $\vec D$, are time independent, since the simplest discussions
of constrained systems assume time-independent constraints.  This assumption can be dropped when the gauge field is quantized along with the Rarita-Schwinger field, but
a more complex system of constraints and constraint brackets will then appear; we defer this extension to a future investigation.

When the constraints are time independent, the classical brackets of Eqs. \eqref{eq:vanishbracks} and \eqref{eq:nonvanishbracks} have the form needed to apply the Faddeev-Popov  \cite{faddeev} method for path integral
quantization. (This has been applied in the free Rarita-Schwinger case by Das and Freedman \cite{dasb} and by Senjanovi\'c \cite{sejn}.) The general formula of \cite{faddeev} for the in to out $S$ matrix element (up to a constant proportionality factor) reads
\begin{align}\label{eq:smatrix}
\langle {\rm out}|S| {\rm in} \rangle  \propto &\int \exp\big(iS(q,p)\big) \prod_t d\mu\big(q(t),p(t)\big)~~~,\cr
d\mu(q,p)=&\prod_a \delta(\chi_a)\delta(\phi_a) (\det [\phi_a,\chi_b])^\xi \prod_i dp_i dq_i~~~,\cr
\end{align}
where $\xi=1$ when all canonical variables are bosonic, and $\xi=-1$ in our case in which all canonical variables
are fermionic, or Grassmann odd.  In applying this formula, we note that since the action $S$ of Eq. \eqref{eq:leftaction}
and the bracket matrix $M$ of Eqs. \eqref{eq:nonvanishbracks}-\eqref{eq:abcdformulas}  are
 independent of $P_{\Psi_0}$ and $P_{\Psi_0^{\dagger}}$, we can immediately integrate out the delta functions in these
two constraints.  Also, since the canonical momentum $\vec P$ is related to $\vec{\Psi}^{\dagger}$ by the constant numerical
transformation of Eqs. \eqref{eq:psidagger} and \eqref{eq:matrixform}, we can take $\vec{\Psi}^{\dagger}$ as the integration variable instead of
$\vec P$, up to an overall proportionality constant.  So we have  the formula
\begin{align}\label{eq:smatrix1}
\langle {\rm out}|S| {\rm in} \rangle\propto & \int \exp\big(i\frac{1}{2}\int d^4x  [-\Psi_{0}^{\dagger} \vec \sigma \cdot \vec D \times \vec {\Psi}
+\vec {\Psi}^{\dagger} \cdot \vec \sigma \times \vec D \Psi_{0}
+\vec{\Psi} ^{\dagger} \cdot \vec D \times \vec \Psi - \vec{\Psi}^{\dagger} \cdot \vec \sigma \times D_{0} \vec{\Psi}]\big) \cr
\times&
\prod_{t,\vec x} d\mu\big(\Psi_0,\Psi_0^{\dagger},\vec \Psi, \vec \Psi^{\dagger}\big)~~~,\cr
\end{align}
with
\begin{equation}\label{eq:measure}
d\mu\big(\Psi_0,\Psi_0^{\dagger},\vec \Psi, \vec \Psi^{\dagger}\big)
=\left(\prod_{c=2}^4 \delta(\chi_c)\delta(\phi_c)\right) (\det [\phi_a,\chi_b])^{-1} d\Psi_0 d\Psi_0^{\dagger} d\vec \Psi d\vec \Psi^{\dagger}~~~,
\end{equation}
with $d\Psi_0$ and $d\Psi_0^{\dagger}$ each a product over the spinor components, and
$d\vec \Psi$ and $d\vec \Psi^{\dagger}$ each a product over the spinor-vector components.

As our next step, we can carry out the integrations over $\Psi_0$ and $\Psi_0^{\dagger}$, using the delta functions $\delta(\phi_2)$ and $\delta(\chi_2)$ .  This leaves the
formula
\begin{align}\label{eq:smatrix2}
\langle {\rm out}|S| {\rm in} \rangle\propto& \int \exp\big(i\frac{1}{2}\int d^4x  [-\vec{\Psi}^{\dagger}
\cdot (\vec B + \vec \sigma \times \vec E) (\vec \sigma \cdot \vec B)^{-1} \vec \sigma \cdot \vec D \times \vec {\Psi}\cr
+&\vec {\Psi}^{\dagger} \cdot \vec \sigma \times \vec D (\vec \sigma \cdot \vec B)^{-1}(\vec B + \vec \sigma \times \vec E) \cdot \vec \Psi
+\vec{\Psi} ^{\dagger} \cdot \vec D \times \vec \Psi - \vec{\Psi}^{\dagger} \cdot \vec \sigma \times D_{0} \vec{\Psi}]\big) \cr
\times&
\prod_{t,\vec x} d\mu\big(\vec \Psi, \vec \Psi^{\dagger}\big)~~~,\cr
\end{align}
with
\begin{equation}\label{eq:measure2}
d\mu\big(\vec \Psi, \vec \Psi^{\dagger}\big)
=\left(\prod_{c=3}^4 \delta(\chi_c)\delta(\phi_c)\right) (\det [\phi_a,\chi_b])^{-1} d\vec \Psi d\vec \Psi^{\dagger}~~~,
\end{equation}
so that the only remaining constraints $\phi_{3,4},\,\chi_{3,4}$ are the ones used in constructing the determinant $\det [\phi_a,\chi_b]$.

We now confront a dilemma:  If we impose the constraint $\chi=0$ coming from $\delta(\chi)$, the first term in the exponent drops out, and we are left
with
\begin{align}\label{eq:smatrix3}
\langle {\rm out}|S| {\rm in} \rangle\propto & \int \exp\big(i\frac{1}{2}\int d^4x
\vec {\Psi}^{\dagger} \cdot[ \vec \sigma \times \vec D (\vec \sigma \cdot \vec B)^{-1}(\vec B + \vec \sigma \times \vec E) \cdot \vec \Psi
+ \vec D \times \vec \Psi - \vec \sigma \times D_{0} \vec{\Psi}]\big) \cr
\times&\prod_{t,\vec x} d\mu\big(\vec \Psi, \vec \Psi^{\dagger}\big)~~~,\cr
\end{align}
in which the coefficient of $\vec {\Psi}^{\dagger}$ in the exponent is $-\vec V$, and so at the stationary phase point of the exponent
with respect to variations in $\vec {\Psi}^{\dagger}$ we get the correct equation of motion.  On the other hand, if we integrate the first
term in the exponent of Eq. \eqref{eq:smatrix3} by parts, it becomes $-(i/2) \int d^4x \chi^{\dagger} (\vec \sigma \cdot \vec B)^{-1}(\vec B + \vec \sigma \times \vec E) \cdot \vec \Psi$,
which is set to zero by the constraint $\chi^{\dagger}=0$ coming from $\delta(\chi^{\dagger})$!  Are we not allowed to integrate by parts here because the
factor $ (\vec \sigma \cdot \vec B)^{-1}(\vec B + \vec \sigma \times \vec E)$ is not continuously differentiable?  This dilemma is avoided by working in
$\Psi_0=0$  gauge, since as we have seen, then $(\vec B + \vec \sigma \times \vec E) \cdot \vec \Psi=0$ and the troublesome term is absent from both the path integral and the equation of motion.   In $\Psi_0=0$  gauge, we end up with the elegant formula
\begin{align}\label{eq:smatrix4}
\langle {\rm out}|S| {\rm in} \rangle\propto & \int \exp\big(i\frac{1}{2}\int d^4x
\vec {\Psi}^{\dagger} \cdot[ \vec D \times \vec \Psi - \vec \sigma \times D_{0} \vec{\Psi}]\big) \cr
\times&\prod_{t,\vec x} d\mu\big(\vec \Psi, \vec \Psi^{\dagger}\big)~~~.\cr
\end{align}
In using  this formula, the customary procedure \cite{fradkin} would be to put the bracket matrix that is the argument of the determinant back into the exponent by introducing bosonic ghost fields $\phi_G$.  But since for $\Psi_0=0$ gauge $\vec L=\vec B + \vec \sigma \times \vec E$,
$\phi_G^{\dagger} N \phi_G$ is only linear in derivatives of $\phi_G$, so one would not get ghost propagators of the standard bosonic form.

\section{Hamiltonian form of the equations and the Dirac bracket}
An alternative way to quantize is to transform the Lagrangian equations to Hamiltonian form, and to take the constraints into
account by replacing the classical brackets by Dirac brackets.  In carrying this out, we will simplify the formulas by
making the gauge choice $A_0=0$ for the non-Abelian gauge fields.  This  gauge choice is always attainable, and leaves a residual non-Abelian
gauge invariance $\vec A \to \vec A + \vec D \Lambda(\vec x)$,with the gauge parameter $\Lambda$ time independent.   The Hamiltonian will then be covariant with respect to this restricted gauge transformation.  For the moment, in discussing the canonical Hamiltonian and bracket formalism,  we will allow $\vec A$ to be time dependent, so that $\vec E \ne 0$.  But when we turn to the Dirac bracket construction corresponding to a constrained Hamiltonian, we will assume a time-independent $\vec A$ as in
the path integral discussion, corresponding in $A_0=0$ gauge to $\vec E=0$. (If we carry along the $A_0$  term in the formulas, which we have done as a check, then time-independent
fields would not require $\vec E=0$.  So this specialization can be avoided at the price of somewhat lengthier equations.)

From the action $S(\Psi_{\mu})=\int dt L(\Psi_{\mu})$  of Eq. \eqref{eq:leftaction} and the canonical momentum of Eq. \eqref{eq:canmom}, we find the canonical Hamiltonian to be
\begin{align}\label{eq:ham}
H=&\int d^3x \partial_0 \vec \Psi \cdot \vec P    - L~~~\cr
=-&\frac{1}{2}\int d^3x  [-\Psi_{0}^{\dagger} \vec \sigma \cdot \vec D \times \vec {\Psi}
+\vec {\Psi}^{\dagger} \cdot \vec \sigma \times \vec D \Psi_{0}
+\vec{\Psi} ^{\dagger} \cdot \vec D \times \vec \Psi ]~~~\cr
=-&\frac{1}{2}\int d^3x  [-\Psi_{0}^{\dagger} \vec \sigma \cdot \vec D \times \vec {\Psi}
+(i\vec P-\vec P \times \vec \sigma) \cdot (\vec \sigma \times \vec D \Psi_{0}
+\vec D \times \vec \Psi) ]~~~,\cr
\end{align}
where in the final line we have used Eq. \eqref{eq:psidagger} to express $\vec{\Psi}^{\dagger}$ in terms of  $\vec P$.

We can now compute the classical brackets of various quantities with $H$.  From
\begin{align}\label{eq:psieqmo}
\frac{d\vec \Psi}{dt}=&[\vec \Psi, H]_C=\frac{1}{2}[i(\vec \sigma \times \vec D \Psi_0+\vec D\times \vec \Psi)-\vec \sigma \times (\vec \sigma \times \vec D \Psi_0+\vec D\times \vec \Psi)]\cr
=&\vec D \Psi_0+ \frac{1}{2} [-\vec \sigma \times (\vec D \times \vec{\Psi}) + i \vec D \times \vec \Psi]~~~,\cr
\end{align}
we obtain the $\vec \Psi$ equation of motion in the form given in Eq. \eqref{eq:d0psi1}.  Similarly,
from the bracket of $\vec P$ with $H$ we find the equation of motion for $\vec{\Psi}^{\dagger}$.
Turning to brackets of the constraints with $H$, starting with $P_{\Psi_0^{\dagger}}$,   we find
\begin{equation}\label{eq:P0dagbrac}
\frac{dP_{\Psi_0^{\dagger}}}{dt}=[P_{\Psi_0^{\dagger}},H]_C=-\frac{1}{2}\chi~~~,
\end{equation}
and so $P_{\Psi_0^{\dagger}}=0$ for all times implies that $\chi=0$.  For the total time derivative
of $\chi$, we have
\begin{equation}\label{eq:chibrac}
\frac{d\chi}{dt}=\frac{\partial \chi}{\partial t}+[\chi,H]_C=\vec \sigma \times g\frac{\partial  \vec A}{\partial t}\cdot \vec{\Psi}+ [\chi,H]_C=-ig \omega~~~,
\end{equation}
and so $\chi=0$ for all times implies that $\omega$ defined in Eq. \eqref{eq:omegatheta2} vanishes.  Since $\omega$ contains
a term proportional to $\Psi_0$, to continue this process by calculating the time derivative of $\omega$, we must obtain
$d\Psi_0/dt$ from a bracket of $\Psi_0$ with $H$ (and similarly for $d\Psi_0^{\dagger}/dt)$.  This requires adding to $H$ a term
\begin{equation}\label{eq:deltah}
\Delta H=-\int d^3x \left[P_{\Psi_0} \frac {d\Psi_0}{dt} + P_{\Psi_0^{\dagger}}\frac{d\Psi_0^{\dagger}}{dt}\right]~~~.
\end{equation}
Requiring $\Delta H$ to be self-adjoint then imposes the requirement
\begin{equation}\label{eq:Padjoint}
P_{\Psi_0}^{\dagger}=- P_{\Psi_0^{\dagger}}~~~,
\end{equation}
which was noted following Eq. \eqref{eq:chidefs}.  We remark again that the chain of successive brackets with $H$ starting from
$P_{\Psi_0^{\dagger}}$ and continuing to $\chi,\omega,...$ leads only to constraints involving $\vec \Psi$ and $\Psi_0$ but never
their adjoints.  The doubling of the set of constraints, which turns the first class constraints into second class ones, comes
from requiring that the adjoint of each fermionic constraint also be a constraint, not from taking successive brackets with $H$.

We are now ready to implement the Dirac bracket procedure.  The basic idea is to change the canonical bracket $[F,G]_C$ to a modified
bracket $[F,G]_D$, which projects $F$ and $G$ onto the subspace obeying the constraints, so that the constraints are built into
the brackets, or after quantization, into the canonical anticommutators.  The constraints  can then be``strongly'' implemented
in the Hamiltonian by setting terms proportional to the constraints to zero.  In doing this we will choose the gauge $\Psi_0=\Psi_0^{\dagger}=0$,
 as we did in the path integral discussion.  Since after integration by parts the second line of Eq. \eqref{eq:ham} takes the form
\begin{equation}\label{eq:ham1}
H=-\frac{1}{2}\int d^3x  [-\Psi_{0}^{\dagger} \chi - \chi^{\dagger}  \Psi_{0}
+\vec{\Psi} ^{\dagger} \cdot \vec D \times \vec \Psi ]~~~,
\end{equation}
this eliminates problems associated with the fact that $\Psi_0,\, \Psi_0^{\dagger}$ have as coefficients
the constraints $\chi^{\dagger},\, \chi$ respectively. Setting constraint terms to zero in Eq. \eqref{eq:ham1}
we see that the constrained Hamiltonian is just
\begin{align}\label{eq:ham2}
H=&-\frac{1}{2}\int d^3x
\vec{\Psi} ^{\dagger} \cdot \vec D \times \vec \Psi \cr
=&-\frac{1}{2}\int d^3x
(i \vec P - \vec P \times \vec \sigma) \cdot \vec D \times \vec \Psi \cr
\end{align}
which coincides with the energy integral computed in Eq. \eqref{eq:graven} from
the stress-energy tensor.

We proceed now to calculate the Dirac bracket for the case when $F=F(\vec \Psi)$ and $G=G(\vec \Psi, \vec \Psi^{\dagger}
)$; the case
when $F=F(\vec \Psi^{\dagger})$ can then be obtained by taking the adjoint, and the case when  $F=F(\vec \Psi, \vec \Psi^{\dagger})$
can be obtained by combining the extra bracket terms from both calculations. When $F$  has no dependence on $\vec \Psi^{\dagger}$,
it has vanishing brackets with the constraints $\phi_a$ of Eq. \eqref{eq:phicon} and nonvanishing brackets with the constraints
$\chi_a$ of Eq. \eqref{eq:chidefs}.  The Dirac bracket then has the form \big(see Eqs. \eqref{a20} and \eqref{a21} for why $M^{-1}$ appears\big)
\begin{equation}\label{eq:dirac1}
[F,G]_D=[F,G]_C-\sum_a\sum_b [F,\chi_a]_C M^{-1}_{ab} [\phi_b,G]~~~,
\end{equation}
where $M_{ab}(\vec x, \vec y) =  [\phi_a(\vec x),\chi_b(\vec y)]_C$ is the matrix defined in
Eqs. \eqref{eq:nonvanishbracks} and \eqref{eq:pstructure}. We recall that this matrix has the form
\begin{equation}\label{eq:pstructure1}
M=\left( \begin{array} {cccc}
 0&-1&0&0 \\
 1&{\cal U}&{\cal S}&{\cal T} \\
 0&{\cal V}&{\cal A}&{\cal B} \\
 0&{\cal W}&{\cal C}&{\cal D} \\
 \end{array} \right)~~~,
\end{equation}
where in the $SU(n)$ gauge field case, each entry in $M$ is a $2n\times 2n $ matrix.  Using the block inversion
method given in Eqs. \eqref{a18} and \eqref{a19}, we find that $M^{-1}$ is given by
\begin{equation}\label{eq:minverse}
M^{-1}=\left( \begin{array} {cccc}
 \Sigma&1&-({\cal S}\cal{F}+{\cal T}{\cal H})&-({\cal S}{\cal G}+{\cal T}\cal{I}) \\
 ~-1~~~&~~~0~~~&~~~0~~~&~~~0~~~ \\
 {\cal F}{\cal V}+{\cal G}{\cal W}&0&{\cal F}&{\cal G} \\
 {\cal H}{\cal V}+{\cal I}{\cal W}&0&{\cal H}&{\cal I} \\
 \end{array} \right)~~~,
\end{equation}
where
\begin{equation}\label{eq:sigmadef}
\Sigma={\cal U}-{\cal S}( {\cal F}{\cal V}+{\cal G}{\cal W})-{\cal T}( {\cal H}{\cal V}+{\cal I}{\cal W} )~~~,
\end{equation}
and where ${\cal F}$, ${\cal G}$, ${\cal H}$, ${\cal I}$ are the elements of the block inversion of the matrix
$N$ of Eq. \eqref{eq:detp},
\begin{equation}
\left(\begin{array} {cc}
  {\cal F}&{\cal G} \\
 {\cal H}&{\cal I} \\
 \end{array} \right)  \left(\begin{array} {cc}
  {\cal A}&{\cal B} \\
 {\cal C}&{\cal D} \\
 \end{array} \right)
=\left(\begin{array} {cc}
  1&0 \\
 0&1 \\
 \end{array} \right)  ~~~.
\end{equation}
Substituting these into Eq. \eqref{eq:dirac1} we find for the Dirac bracket a lengthy expression, which
simplifies considerably after noting that $[F(\vec \Psi),\chi_1]_C=[F(\vec \Psi), -P_{\Psi_0}]_C=0$ and
$[\phi_1,G(\vec \Psi, \vec{\Psi}^{\dagger})]_C=[P_{\Psi_0^{\dagger}},G(\vec \Psi, \vec{\Psi}^{\dagger})]_C=0$,
leaving the relatively simple formula
\begin{align}\label{eq:simplebrac}
[F(\vec \Psi),G(\vec \Psi, \vec{\Psi}^{\dagger})]_D=& [F(\vec \Psi),G(\vec \Psi, \vec{\Psi}^{\dagger})]_C   \cr
-& [F(\vec \Psi),\chi_3]_C \Big({\cal F}\, [\phi_3,G(\vec \Psi, \vec{\Psi}^{\dagger})]_C + {\cal G} \,[\phi_4,G(\vec \Psi, \vec{\Psi}^{\dagger})]_C\Big)   \cr
-& [F(\vec \Psi),\chi_4]_C \Big({\cal H}\, [\phi_3,G(\vec \Psi, \vec{\Psi}^{\dagger})]_C + {\cal I}\, [\phi_4,G(\vec \Psi, \vec{\Psi}^{\dagger})]_C\Big) ~~~.  \cr
\end{align}
We note that just as in the path integral derivation, only the matrix $N$ enters, in this case through its inverse, rather than the full
matrix of constraint brackets $M$.  The final step is to evaluate the inverse block matrix elements ${\cal F},\,{\cal G},\,{\cal H},\,{\cal I}$ from
the expressions for ${\cal A},\,{\cal B},\,{\cal C},\,{\cal D}$, again by using the block inversion formulas of Eqs. \eqref{a18} and \eqref{a19}.
Let us define the Green's function ${\cal D}^{-1}(\vec x-\vec y)$
by
\begin{equation}\label{eq:green}
\big( i(\vec L_{\vec x})^2 + \vec \sigma \cdot \vec L_{\vec x} \times \vec L_{\vec x} \big){\cal D}^{-1}(\vec x-\vec y)=\delta^3(\vec x-\vec y)~~~,
\end{equation}
and a second Green's function ${\cal Z}(\vec x - \vec y)$ by
\begin{align}\label{eq:greenz}
{\cal Z}(\vec x - \vec y)=&{\cal A} - {\cal B} {\cal D}^{-1} {\cal C}\cr
=&-2ig \vec  \sigma \cdot \vec B \delta^3(\vec x-\vec y) - 4 \vec D_{\vec x} \cdot \vec L_{\vec x} {\cal D}^{-1}(\vec x-\vec y) \vec{L}_{\vec y} \cdot \overleftarrow{D}_{\vec y}~~~.\cr
\end{align}
where we recall that for $\Psi_0=0$ gauge $\vec L=\vec B+\vec \sigma \times \vec E$.
Then the needed inverse block matrices are
\begin{align}\label{eq:inverseblock}
{\cal F}=&{\cal Z}^{-1}~~~,\cr
{\cal G}=&-{\cal Z}^{-1}{\cal B}{\cal D}^{-1}~~~,\cr
{\cal H}=&-{\cal D}^{-1}{\cal C}{\cal Z}^{-1}~~~,\cr
{\cal I}=&{\cal D}^{-1} + {\cal D}^{-1} {\cal C} {\cal Z}
^{-1} {\cal B} {\cal D}^{-1}~~~.\cr
\end{align}

We wish now to apply the Dirac bracket formula to the cases (i)  $F(\vec \Psi)=\vec \Psi$ and $G(\vec \Psi, \vec{\Psi}^{\dagger})=\vec{\Psi}^{\dagger}$, and
(ii) $F(\vec \Psi)=\vec \Psi$ and $G(\vec \Psi, \vec{\Psi}^{\dagger})=H$, with $H$ the constrained Hamiltonian of Eq. \eqref{eq:ham2}.
The following canonical brackets are needed for this:
\begin{align}\label{eq:neededbracs}
[\vec \Psi(\vec x),\chi_3(\vec y)]_C=&2 \vec D_{\vec x} \delta^3(\vec x-\vec y)~~~,\cr
[\vec \Psi(\vec x),\chi_4(\vec y)]_C=&(i \vec L_{\vec x}- \vec \sigma \times \vec L_{\vec x})\delta^3(\vec x-\vec y)~~~,\cr
[\phi_3(\vec x),\vec{\Psi}^{\dagger}(\vec y)]_C=&2 \vec D_{\vec x} \delta^3(\vec x-\vec y)= -2 \delta^3(\vec x-\vec y) \overleftarrow{D}_{\vec y}~~~,\cr
[\phi_4(\vec x),\vec{\Psi}^{\dagger}(\vec y)]_C=&-(i \vec L_{\vec x}-  \vec L_{\vec x} \times  \vec \sigma) \delta^3(\vec x-\vec y)
=\delta^3(\vec x-\vec y)(i \overleftarrow{L}_{\vec y}- \overleftarrow{L}_{\vec y} \times \vec \sigma )~~~,\cr
[\phi_3(\vec x),H]_C=& ig\vec B(\vec x) \cdot \vec \Psi(\vec x)~~~,\cr
[\phi_4(\vec x),H]_C=&\frac{1}{2} (i\vec L_{\vec x}- \vec L_{\vec x} \times \vec \sigma) \times \vec D_{\vec x} \cdot \vec \Psi(\vec x)~~~. \cr
\end{align}
Additionally, for case (i) we need the canonical bracket
\begin{align}\label{eq:psipsibarbrac}
[\Psi_i(\vec x),\Psi^{\dagger}_j(\vec y)]_C= &[\Psi_i(\vec x),iP_j(\vec y)-\epsilon_{jkl}P_k(\vec y)\sigma_l]_C\cr
= &-i (\delta_{ij}+i\epsilon_{jil}\sigma_l)\delta^3(\vec x-\vec y)=-i \sigma_j\sigma_i \delta^3(\vec x-\vec y)=-2i \Big(\delta_{ij}-\frac{1}{2}\sigma_i\sigma_j\Big)\delta^3(\vec x-\vec y)  ~~~,\cr
\end{align}
and for case (ii) we need the canonical bracket
\begin{equation}\label{eq:psiHbrac}
[\Psi_i(\vec x),H]_C=\frac{1}{2}\Big(i \vec D_{\vec x} \times \vec \Psi(\vec x) - \vec \sigma \times \big(\vec D_{\vec x} \times \vec \Psi(\vec x)\big)\Big)_i~~~.
\end{equation}

Up to this point, we have not specialized  $\vec L$  so as to make it easy to ascertain what the formulas become when gauge fixing is omitted (as in
\cite{johnson} and \cite{velo}). When $\vec L=0$, the matrix $N$ degenerates to its upper left element ${\cal A}$, which is a local function of $\vec x$ and so
is algebraically invertible.  For the Dirac bracket of $\vec \Psi_i(\vec x)$ with $\vec \Psi^{\dagger}_j(\vec y)$ we then find
\begin{align}\label{eq:velobrac}
[\Psi_i(\vec x),\Psi^{\dagger}_j(\vec y)]_D= &[\Psi_i(\vec x),\Psi^{\dagger}_j(\vec y)]_C-\int d^3w d^3z [\Psi_i(\vec x),\chi_3(\vec w)]_C {\cal Z}^{-1}(\vec w-\vec z) [\phi_3(\vec z),\Psi^{\dagger}_j(\vec y)]_C\cr
=&-2i\Big[ \Big(\delta_{ij}-\frac{1}{2}\sigma_i\sigma_j\Big)\delta^3(\vec x-\vec y)-D_{\vec x \,i}\frac{\delta^3(\vec x-\vec y)}{g \vec \sigma \cdot \vec B(\vec x)}\overleftarrow{D}_{\vec y\,j}\Big]\cr
=&-2i\langle \vec x|\Big[ \Big(\delta_{ij}-\frac{1}{2}\sigma_i\sigma_j\Big)1+\Pi_i\frac{1}{g \vec \sigma \cdot \vec B}\Pi_j\Big]|\vec y \rangle~~~,\cr
\end{align}
where in the final line we have written $iD_{\vec x\, i}=\Pi_i$ to relate to the abstract operator notation of Velo and Zwanziger \cite{velo}.  Multiplying the final line by $i$ to
convert the Dirac bracket to an anticommutator, and by a factor $1/2$ reflecting our different field normalization, Eq. \eqref{eq:velobrac} becomes  the expression for the
anticommutator given in the zero mass limit of Eq. (4.12) of \cite{velo}.  Using identities in Appendix A, one can verify (as in Appendix C of \cite{velo}) that
\begin{equation}\label{eq:chiproj}
(\vec \sigma \times \vec D_{\vec x})_i \Big[ \Big(\delta_{ij}-\frac{1}{2}\sigma_i\sigma_j\Big)\delta^3(\vec x-\vec y)-D_{\vec x\,i}\frac{\delta^3(\vec x-\vec y)}{g \vec \sigma \cdot \vec B(\vec x)}\overleftarrow{D}_{\vec y\,j}\Big]=0~~~,
\end{equation}
that is, the constraint $\chi$ is explicitly projected to zero.

Now setting $\vec L=\vec B + \vec \sigma \times \vec E$, so that the $\omega$ constraint implies $\Psi_0=\Psi_0^{\dagger}=0$,  we find for the Dirac bracket of $\Psi_i(\vec x)$  with the constrained Hamiltonian
\begin{align}\label{eq:psihambrac}
\frac{d\vec \Psi(\vec x)}{dt}=& [\vec \Psi(\vec x),H]
_D=\frac{1}{2}[i\vec D_{\vec x} \times \vec \Psi(\vec x)-\vec \sigma \times \big(\vec D_{\vec x} \times \vec \Psi(\vec x) \big)]\cr
-&\int d^3y\Big\{ 2 \vec D_{\vec x}\Big[{\cal F}(\vec x-\vec y) i g \vec B(\vec y) \cdot \vec \Psi(\vec y)\cr
+&{\cal G}(\vec x-\vec y)
 \frac{1}{2}\big(i \vec L_{\vec y}-\vec L_{\vec y} \times \vec \sigma\big) \times \vec D_{\vec y} \cdot \vec \Psi(\vec y)\Big]  \cr
+&(i \vec L_{\vec x} - \vec \sigma \times \vec L_{\vec x}) \Big[{\cal H}(\vec x-\vec y) i g \vec B(\vec y) \cdot \vec \Psi(\vec y)\cr
+&{\cal I}(\vec x-\vec y) \frac{1}{2}\big(i \vec L_{\vec y}-\vec L_{\vec y} \times \vec \sigma\big) \times \vec D_{\vec y} \cdot \vec \Psi(\vec y)\Big]\Big\}~~~.
\end{align}
The first line of this equation is the unconstrained equation of motion in the form of Eq. \eqref{eq:d0psi1} (when $A_0=\Psi_0=0$), while the remaining terms guarantee that
\begin{align}\label{eq:chivanish}
\frac{d\phi_3}{dt}=&\frac{d\chi}{dt}= \frac{d(\vec \sigma \times \vec D \cdot \vec \Psi)}{dt}=\sigma \times \vec D \cdot \frac{d\vec \Psi}{dt}=0~~~,\cr
\frac{d\phi_4}{dt}=&d\frac{\vec L \cdot \vec \Psi}{dt}=\vec L \cdot  \frac{d\vec \Psi}{dt}=0~~~,\cr
\end{align}
where we have used the fact that we are assuming that $\vec D$ and  $\vec L$ are time independent.  This restriction can be avoided by treating the gauge fields
as dynamical variables, taking into account their own constraint structure, and noting that the radiation gauge fixing constraint $\vec \nabla \cdot \vec P_{\vec A}=0$,
with $\vec P_{\vec A}$ the canonical momentum conjugate to $\vec A$, has nonvanishing fermionic  brackets with all Rarita-Schwinger constraints involving $\vec D=\vec \nabla + g \vec A$.  This requires an extension of the Dirac bracket construction to take the new, Grassmann-odd, brackets into account, and the extended Dirac bracket structure will then
obey Eq. \eqref{eq:chivanish} without requiring the assumption of a time independent $\vec A$ and $\vec L$.

In the next section, where we show positivity of the Dirac anticommutator, we will use covariant radiation gauge instead of $\Psi_0=0$ gauge.
Setting $\vec L = \vec D$ and putting everything together, we find for the Dirac bracket of $\Psi_i(\vec x)$ with $\Psi_j^{\dagger}(\vec y)$,
\begin{align}\label{eq:psipsidagbrac}
[\Psi_i(\vec x),\Psi_j^{\dagger}(\vec y)]_D=&-2i \Big(\delta_{ij}-\frac{1}{2}\sigma_i\sigma_j\Big)\delta^3(\vec x-\vec y) \cr
+& 4\vec  D_{\vec x\, i}{\cal F}(\vec x-\vec y) \overleftarrow{D}_{\vec y\,j} -2  D_{\vec x\, i} {\cal G}(\vec x-\vec y)
(i\overleftarrow{D}_{\vec y}- \overleftarrow{D}_{\vec y} \times \vec \sigma)_j\cr
+&2(i\vec  D_{\vec x}- \vec \sigma \times \vec  D_{\vec x})_i {\cal H}(\vec x-\vec y) \overleftarrow{D}_{\vec y\, j}-(i\vec  D_{\vec x}
- \vec \sigma \times \vec  D_{\vec x})_i{\cal I}(\vec x -\vec y)  (i\overleftarrow{D}_{\vec y}- \overleftarrow{D}_{\vec y} \times \vec \sigma )_j~~~,\cr
\end{align}
which gives the generalization of Eq. \eqref{eq:velobrac} to the case when a covariant gauge fixing constraint is imposed.

\section{Quantization of the anticommutator derived from the Dirac bracket and positivity  in covariant radiation gauge}

Given the Dirac bracket, the next step is to quantize, by multiplying all Dirac brackets by $i$ and then reinterpreting them as anticommutators or commutators of operators.  In the
case considered here, this can be done in a constructive way, as follows.  First let us replace the set of $2n$ component column vector constraints $\phi_a$ and $2n$ component row vector constraints $\chi_a$ by the set of $4n$ scalars given by their individual matrix elements.  Moreover, since the $\chi_a$ are the adjoints of the $\phi_a$, we can take linear combinations to make
all of these scalars self-adjoint.  Labeling the set of self-adjoint scalar constraints by $\Phi_a$, the Dirac bracket construction for the bracket of $F$ with $G$ reads
\begin{align}\label{eq:fullbrac}
[F,G]_D=&[F,G]_C-\sum_a\sum_b [F,\Phi_a]_C T^{-1}_{ab} [\Phi_b,G]_C~~~,\cr
T_{ab}=&[\Phi_a,\Phi_b]_C~~~,  \cr
\end{align}
with the matrix $T$ real.

We now observe that since the $\Phi_a$ are all {\it linear} in the scalar components of $\vec \Psi$ and  $ \vec \Psi^{\dagger}$, if we make the replacement $i[~,~]_C \to \{~,~\}_C$, with $\{~,~\}
$ the anticommutatior, and replace all Grassmann variables $\vec \Psi$ and $\vec \Psi^{\dagger}$ with operator variables having
the standard canonical anticommutators, then since there is no other operator structure the same real matrix $T_{ab}$ will be obtained.  Moreover,
if $F$ and $G$ are both linear in the scalar components of $\vec \Psi$ and  $\vec \Psi^{\dagger}$, the Grassmann bracket  $i[F,G]_C$ formed from scalar components of $F$ and $G$ will agree  with the canonical anticommutator  $i\{F,G\}_C$ formed from the corresponding operator scalar components, and will be a $c$-number.   Thus, for linear
$F$ and $G$ we can define a ``Dirac anticommutator" $\{F,G\}_D$  by
\begin{align}\label{eq:anticommbrac}
\{F,G\}_D=&\{F,G\}_C-\sum_a\sum_b \{F,\Phi_a\}_C T^{-1}_{ab} \{\Phi_b,G\}_C~~~,\cr
T_{ab}=&\{\Phi_a,\Phi_b\}_C~~~.\cr
\end{align}
When one or both of $F$ and $G$ is bilinear, the Grassmann bracket $i[F,G]_C$ formed from the scalar components of $F$ and $G$ will agree with the
canonical commutator formed from the corresponding operator scalar components, and we can  define a ``Dirac commutator'' by  a formula analogous to Eq. \eqref{eq:anticommbrac} in which each anticommutator with at least one bilinear argument is replaced by a commutator.  In this way  we get
a mapping of classical brackets into quantum anticommutators and commutators, that inherits  the algebraic properties of the Dirac bracket, including the
chain rule, with the Jacobi identities for odd and even Grassmann variables mapping to the corresponding   anticommutator and commutator Jacobi identities.

To complete this correspondence, we must show that the Dirac anticommutator of $\Psi_i^{\alpha\,u}$ and $\Psi_j^{\dagger \, \beta \,v}$ (with $\alpha =1,2,~\beta=1,2$ the spin indices, $u=1,...,n,~v=1,...,n$ the internal symmetry indices, and $i=1,2,3,~j=1,2,3$ the spatial vector
 indices) has the expected positivity properties of an
operator anticommutator, by showing that for an arbitrary set of complex functions $A_i^{\alpha\,u}(\vec x)$, we have
\begin{equation}\label{eq:pos1}
\int d^3x d^3y A_i^{\alpha\,u}(\vec x) A_j^{*\beta\,v}(\vec y) \{\Psi_i^{\alpha\,u}(\vec x),\Psi_j^{\dagger \, \beta\,v}(\vec y)\}_D \geq 0~~~.
\end{equation}
We demonstrate this in several steps, in covariant radiation gauge.  First we examine the conditions for positivity of the canonical anticommutator
and Poisson bracket,
\begin{equation}\label{eq:pos2}
\int d^3x d^3y A_i^{\alpha\,u}(\vec x) A_j^{*\beta\,v}(\vec y) \{\Psi_i^{\alpha\,u}(\vec x),\Psi_j^{\dagger \, \beta\,v }(\vec y)\}_C= \int d^3x d^3y  A_i^{\alpha\,u}(\vec x) A_j^{*\beta\,v}(\vec y) i[\Psi_i^{\alpha\,u}(\vec x),\Psi_j^{\dagger \, \beta\,v }(\vec y)]_C~~~.
\end{equation}\label{eq:pos3a}
From $\Psi_j^{\dagger \, \beta \,v}= i  P_j^{\beta\,v}-\epsilon_{jkl} P_k^{\delta\,v} \sigma_l^{\delta \beta}$,
we find that
\begin{align}\label{eq:pos3}
[\Psi_i^{\alpha\,u}(\vec x),\Psi_j^{\dagger \, \beta \,v}(\vec y)]_C=&-i\big(\delta_{ij}\delta^{\alpha\beta}+i\epsilon_{jik} \sigma_k^{\alpha \beta}\big)\delta^{uv}\delta^3(\vec x-\vec y)\cr
 =& -i (\sigma_j\sigma_i)^{\alpha\beta}\delta^{uv}\delta^3(\vec x-\vec y)
= -2i (\delta_{ij}-\frac{1}{2}\sigma_i\sigma_j)^{\alpha\beta}\delta^{uv}\delta^3(\vec x-\vec y)~~~.\cr
\end{align}
Multiplying by $i/2$, and writing $A_i^{\alpha\,u}=R_i^{\alpha\,u}+iI_i^{\alpha\,u},\, i=1,2,3,\,\alpha=1,2,\,u=1,...,n$, with $R$ and $I$ real, the right hand side of  Eq. \eqref{eq:pos2} evaluates to \big(we suppress the internal symmetry index $u$ from here on, so $(R_i^{\alpha})^2$ means
$\sum_{u=1}^n(R_i^{\alpha\,u})^2$ , etc.\big)
\begin{equation}\label{eq:pos4}
\sum_{i=1}^3\sum_{\alpha=1}^2  \big( (R_i^{\alpha})^2+(I_i^{\alpha})^2\big) - \frac{1}{2}\big((R_2^1-I_1^1+I_3^2)^2+(R_1^1+I_2^1-R_3^2)^2+(R_2^2+I_1^2+I_3^1)^2+(R_1^2-I_2^2+R_3^1)^2\big)~~~.
\end{equation}
If all three components $A_i^{\alpha},\, i=1,...,3$ are present, the expression in Eq. \eqref{eq:pos4} is {\it not} positive semidefinite.  But when only two of the three components are present, as a result of application of a constraint, then each of the four squared terms on the right hand
side of Eq. \eqref{eq:pos4} contains only two terms, and so the expression in Eq. \eqref{eq:pos4} is positive semidefinite by virtue of the inequality
\begin{equation}\label{eq:posineq}
X^2+Y^2-\frac{1}{2}(X\pm Y)^2= \frac{1}{2}(X\mp Y)^2 \geq 0~~~.
\end{equation}
Another way of seeing this, noted by both Velo and Zwanziger \cite{velo} and Allcock and Hall \cite{allcock}, is that because $\sum_{i=1}^3 \sigma_i \sigma_i =3$, the
expression $W_{ij}=\delta_{ij}- \frac{1}{2} \sigma_i\sigma_j$ is not a projector.  But when one component of $\vec \sigma$, say $\sigma_3$,  is replaced by 0, so that one has  $\sum_{i=1}^3 \sigma_i \sigma_i =\sum_{i=1}^2 \sigma_i \sigma_i =2$,
then
\begin{equation}\label{eq:wproj}
\sum_l W_{il}W_{lj}=\delta_{ij}- 2 \frac{1}{2}\sigma_i\sigma_j+\frac{1}{4} \sigma_i \sum_{l=1}^2 \sigma_l \sigma_l \sigma_j = \delta_{ij}-\frac{1}{2}\sigma_i\sigma_j= W_{ij}~~~,
\end{equation}
and $W_{ij}$ is a projector and hence is positive semidefinite.  So we anticipate that proving positivity will require projection of
Eq. \eqref{eq:pos3} into a subspace obeying  at least one constraint on $\vec \Psi$.

The next step is to use the  property that the Dirac bracket of linear quantities $F$ and $G$ reduces to
the canonical bracket of their projections into the subspace obeying the constraints,
when (as is the case here) all constraints are second class, that is they all appear in the Dirac bracket \cite{hanson}.
Referring to Eq. \eqref{eq:fullbrac}, let us define
\begin{align}\label{eq:fproj}
\tilde F=&F-\sum_a \sum_b [F,\Phi_a]_C T^{-1}_{ab} \Phi_b~~~,\cr
\tilde G=&G-\sum_a \sum_b [G,\Phi_a]_C T^{-1}_{ab} \Phi_b~~~,\cr
 \end{align}
so that
\begin{align}\label{eq:tildebrac}
[\tilde F,\Phi_c]_C=&[F,\Phi_c]_C- \sum_a\sum_b [F,\Phi_a]_C T^{-1}_{ab} [\Phi_b,\Phi_c]_C  ~~~\cr
=&[F,\Phi_c]_C- \sum_a\sum_b [F,\Phi_a]_C T^{-1}_{ab}T_{bc}  ~~~\cr
=&[F,\Phi_c]_C- \sum_a [F,\Phi_a]_C \delta_{ac}=0 ~~~,\cr
\end{align}
and similarly for $\tilde G$.  As a result of this relation, which holds when the canonical brackets
are simply numbers (as in the case here where $\Phi_c$ and $F,\,G$ are linear),
together with symmetry  of the canonical bracket $[\tilde G, \Phi_c]_C=[\Phi_c, \tilde G]_C$, we see that
\begin{equation}\label{eq:tildebrac1}
[ F, G]_D= [\tilde F,\tilde G]_C~~~.
\end{equation}
These properties of Eqs. \eqref{eq:fproj}--\eqref{eq:tildebrac1} carry over when we replace Grassmann numbers with operators, and  classical brackets with anticommutators, since
in the linear case all anticommutators of linear quantities are c-numbers that commute with the operators, and since the
anticommutator is symmetric.  Thus we have
\begin{equation}\label{eq:tildebrac3}
  \{ \Psi_i^{\alpha}(\vec x), \Psi_j^{\dagger \, \beta}(\vec y\}_D=   \{\tilde \Psi_i^{\alpha}(\vec x),\tilde \Psi_j^{\dagger \, \beta}(\vec y)\}_C~~~.
\end{equation}

To further study the properties of $\tilde \Psi_i(\vec x)$ and $\tilde
\Psi_j^{\dagger}(\vec y)$  (with spinor indices suppressed), let us now return to our
original labeling of the constraints by $\phi_a$ and $\chi_a$ as in Eq. \eqref{eq:simplebrac}, so that we have
in the Dirac bracket formalism
\begin{equation}\label{eq:originaltilde}
\tilde  \Psi_i(\vec x)=\Psi_i(\vec x)-\sum_a \sum_b [ \Psi_i(\vec x),\chi_a]_C M_{ab}^{-1} \phi_b~~~,
\end{equation}
and a similar equation (with the roles of $\phi_a$ and $\chi_a$ interchanged) for $\tilde \Psi_j^{\dagger}(\vec y)$, with $a,b$ summed from 3 to 4.  We now note two important
properties of this equation.  The first is that it is invariant under replacement of the constraints $\chi_a$ by
any linear combination $\chi_a^{\prime}= \chi_b K_{ba}$, with the matrix $K$ nonsingular, since the factors
$K$ and $K^{-1}$ cancel between $\chi_a^{\prime}$ and $M_{ab}^{\prime\,-1}$.  (More generally, the Dirac bracket is invariant under
replacement of the constraints by any nonsingular linear combination of the constraints, reflecting the fact that the
Dirac bracket is a projector onto the subspace obeying the constraints, and this subspace is invariant
under replacement of the constraints by any nonsingular linear combination of the constraints.) The second is that
if we act on $\tilde \Psi_i(\vec x)$ with either $D_{\vec x\,i}$ or $(\vec \sigma \times D_{\vec x})_i$, we get zero.
For example, recalling that in covariant radiation gauge $D_{\vec x\,i} \Psi_i(\vec x)=\phi_4(\vec x)$, we have (with spatial variable labels
$\vec x$ suppressed)
\begin{equation}\label{eq:dgives0}
D_i\tilde  \Psi_i=\phi_4 -\sum_a \sum_b [ \phi_4,\chi_a]_C M_{ab}^{-1} \phi_b
=\phi_4 -\sum_a \sum_b M_{4a}M_{ab}^{-1} \phi_b=\phi_4 -\sum_b \delta_{4b} \phi_b=0~~~,
\end{equation}
and similarly for  $(\vec \sigma \times D_{\vec x})_i$,  with $\phi_4$ replaced by $\phi_3$.

Let us now write $\tilde  \Psi_i(\vec x)$ as a projector $R_{ij}(\vec x,\vec y)$ acting on $\Psi_j(\vec y)$,  giving
after an integration by parts on $\vec y$,
\begin{align}\label{eq:projectordef}
\tilde  \Psi_i(\vec x)=&\int d^3y R_{ij}(\vec x,\vec y) \Psi_j(\vec y)~~~,\cr
R_{ij}(\vec x,\vec y)=&\delta_{ij}\delta^3(\vec x-\vec y)+ \sum_a \sum_b \int d^3z [ \Psi_i(\vec x),\chi_a(\vec z)]_C M_{ab}^{-1}(\vec z,\vec y)  \overleftarrow{\eta}_{b\,j}(\vec y)~~~,\cr
\end{align}
with
\begin{equation}\label{eq:etadef}
\overleftarrow{\eta}_{3\,j}(\vec y)=(\vec \sigma \times \overleftarrow{D}_{\vec y})_j~~~,~~
\overleftarrow{\eta}_{4\,j}(\vec y)=\overleftarrow{D}_{\vec y\,j}~~~.
\end{equation}
By virtue of Eq. \eqref{eq:dgives0} and its analog for $\vec \sigma \times \vec D$, we have
\begin{align}\label{eq:dgives01}
D_{\vec x\,i} R_{ij}(\vec x,\vec y)=&0~~~,\cr
(\vec \sigma \times \vec D_{\vec x})_i R_{ij}(\vec x,\vec y)=&0~~~,\cr
\end{align}
and by the reasoning of Eqs. \eqref{eq:chisimp}--\eqref{eq:sigmaconstraint} we also have
(assuming $\vec \sigma \cdot \vec D$ is invertible)
\begin{equation}\label{eq:sigmagives0}
\sigma_i R_{ij}(\vec x,\vec y)=0~~~.
\end{equation}
Next let us focus on the bracket $[ \Psi_i(\vec x),\chi_a(\vec z)]_C $ appearing as the first factor inside the sum.
Setting $\vec L=\vec D$  in Eq. \eqref{eq:neededbracs} we have
\begin{align}\label{eq:neededbracs1}
[\vec \Psi(\vec x),\chi_3(\vec z)]_C=&2 \vec D_{\vec x} \delta^3(\vec x-\vec z)~~~,\cr
[\vec \Psi(\vec x),\chi_4(\vec z)]_C=&(i \vec D_{\vec x}- \vec \sigma \times \vec D_{\vec x})\delta^3(\vec x-\vec z)~~~.\cr
\end{align}
Using the invariance of $\tilde \Psi_i$, or equivalently of $R_{ij}$, under replacement of $\chi_3,\,\chi_4$ by any nondegenerate
linear combination of  $\chi_3,\,\chi_4$,  let us choose the new combinations so that
\begin{align}\label{eq:neededbracs2}
[\vec \Psi (\vec x),\chi_3(\vec z)]_C=& (\vec \sigma \times \vec D_{\vec x})\delta^3(\vec x-\vec z)=\vec \eta_3(\vec x)\delta^3(\vec x-\vec z)~~~,\cr
[\vec \Psi(\vec x),\chi_4(\vec z)]_C=& \vec D_{\vec x}\delta^3(\vec x-\vec y)=\vec \eta_4(\vec x)\delta^3(\vec x-\vec z) ~~~.\cr
\end{align}
Substituting this into Eq. \eqref{eq:projectordef}, we get the symmetric  expression
\begin{equation}\label{eq:projectsym}
R_{ij}(\vec x,\vec y)=\delta_{ij}\delta^3(\vec x-\vec y)+ \sum_a \sum_b \int d^3z \vec \eta_{a\,i}(\vec x)  M_{ab}^{-1}(\vec x,\vec y)  \overleftarrow{\eta}_{b\,j}(\vec y)~~~.
\end{equation}
By virtue of this symmetry, the projector $R_{ij}$ is annihilated by the constraints $\overleftarrow{D}_{\vec y\,j}$ and
$(\vec \sigma \times \overleftarrow{D}_{\vec y})_j$ acting from the right, which in turn implies that
in addition to  Eq. \eqref{eq:sigmagives0} we also have
\begin{equation}\label{eq:sigmagives01}
 R_{ij}(\vec x,\vec y)\sigma_j=0~~~.
\end{equation}
An explicit construction of $R_{ij}(\vec x,\vec y)$ and verification of Eqs. \eqref{eq:sigmagives0}    and  \eqref{eq:sigmagives01}   is given in Appendix C.

Returning now to Eqs. \eqref{eq:pos1} and \eqref{eq:tildebrac3}, writing $\tilde \Psi_i^{\alpha}$ and $\tilde \Psi^{\dagger\,\beta}_j$ in terms of  projectors acting
on $\Psi_i^{\alpha}$ and $\Psi^{\dagger\,\beta}_j$, we have  (using $\sigma_m^{\epsilon \delta}=\sigma_m^{*\delta \epsilon}$, and continuing
to suppress internal symmetry indices $u,v$, which are contracted in the same pattern as the spatial vector and spin indices)
\begin{align}\label{eq:tildebrac4}
&\int d^3x  \int d^3y A_i^{\alpha}(\vec x) A_j^{*\, \beta}(\vec y) \{\Psi_i^{\alpha}(\vec x), \Psi_j^{\dagger\, \beta}(\vec y)\}_D \cr= &\int d^3x \int d^3y A_i^{\alpha}(\vec x)A_j^{*\,\beta}(\vec y)  \{\tilde \Psi_i^{\alpha}(\vec x),\tilde \Psi_j^{\dagger\,\beta}(\vec y)\}_C\cr= &
\int d^3x \int d^3y A_i^{\alpha}(\vec x)A_j^{*\beta}(\vec y) \int d^3z \int d^3w \,R_{il}^{\alpha\gamma}(\vec x,\vec z)\{\Psi_l^{\gamma}(\vec z), \Psi_m^{\dagger\,\delta}(\vec w)\}_C R^{*\beta\delta}_{jm}(\vec y,\vec w)\cr
=&\int d^3x \int d^3y A_i^{\alpha}(\vec x)A_j^{*\beta}(\vec y) \int d^3z \int d^3w\, R_{il}^{\alpha\gamma}(\vec x,\vec z)2\left(\delta_{lm}\delta^{\gamma\delta}-\frac{1}{2}\sigma_{l}^{\gamma\epsilon}\sigma_m^{*\delta\epsilon}\right)\delta^3(\vec z-\vec w) R^{*\beta\delta}_{jm}(\vec y,\vec w)\cr
=&2\int d^3z \Big[\int d^3x  A_i^{\alpha} (\vec x)R_{il}^{\alpha\gamma}(\vec x,\vec z)\Big]
\Big[ \int d^3y A_j^{\beta}(\vec y) R^{\beta\gamma}_{jl}(\vec y,\vec z)\Big]^*~~~,\cr
\end{align}
which is  positive semidefinite.

We conclude that the anticommutator of $\vec \Psi$ with $\vec \Psi^{\dagger}$ is positive semidefinite in covariant radiation gauge.  The symmetry
of the $\phi_{3,4}$ and $\chi_{3,4}$ constraints in this gauge is essential to reaching this conclusion; if gauge fixing were omitted,
or if another gauge were chosen, this symmetry would not be present and we could not deduce positivity in a similar fashion.

\section{Lorentz covariance of covariant radiation and $\Psi_0=0$  gauge and Lorentz invariance of the Dirac bracket}

We study next the behavior of covariant radiation gauge and the Dirac bracket under Lorentz boosts.  The Rarita-Schwinger field
$\psi_{\mu}^{\alpha}$ and its left-handed chiral projection $\Psi_{\mu}^{\alpha}$ both have a four-vector index $\mu$ and
a spinor index $\alpha$.  Under an infinitesimal Lorentz transformation, the transformations acting on these two types of indices
are additive, and so can be considered separately.  The spinor indices are transformed as in the usual spin $\frac{1}{2}$ Dirac
equation by a matrix constructed from the Dirac gamma matrices, which commutes with $D_{\mu}$.  Hence the spinor index  transformation leaves
the covariant radiation gauge condition $\vec D \cdot \vec \Psi$ invariant.

This leaves the transformation on the vector index to be considered, and this is a direct analog of the Lorentz transformation of radiation
gauge in quantum electrodynamics \cite{zumino}.  Since the radiation gauge condition is invariant under spatial rotations, we
only have to consider a Lorentz boost,
\begin{align}\label{eq:boost1}
\vec x \to &\vec x^{\,\prime}= \vec x + \vec v t~~~, \cr
x^0=&t \to  t^{\prime} = t+ \vec v \cdot \vec x~~~.\cr
\end{align}
Under this boost, the field $\vec \Psi$ transforms as
\begin{equation} \label{eq:boost2}
\vec \Psi \to \vec \Psi^{\prime} = \vec \Psi + \vec v \Psi^0~~~.
\end{equation}
For an observer in the boosted frame, covariant radiation gauge would be
$\vec D_{\vec x^{\prime}} \cdot \vec \Psi^{\prime}=0$, with
$\vec D_{\vec x^{\prime}}= \vec \nabla_{\vec x^{\prime}}+ g \vec A^{\prime}$, where $\vec A^{\prime}=\vec A + O(\vec v)$.
Applying this to $\vec \Psi^{\prime}(\vec x^{\prime},t^{\prime})$
and using the covariant radiation gauge condition in the initial frame, we get
\begin{equation} \label{eq:vecd1}
\vec D_{\vec x^{\prime}}\cdot  \vec \Psi^{\prime}
= v_j \Sigma_j(\vec x,t)~~~,
\end{equation}
with $\Sigma_j(\vec x,t)$ a local polynomial in $\vec \Psi,~\Psi_0$ and the gauge fields,
where we have dropped primes on the right hand side since there is an explicit factor of $\vec v$.
So in the boosted frame $\vec \Psi^{\prime}$ does not obey the covariant radiation gauge condition, but this
can be restored by making a gauge transformation
\begin{equation}\label{eq:gaugerestore}
\vec \Psi^{\prime} \to \vec \Psi^{\prime} -\vec D
(\vec D^2)^{-1} v_j \Sigma_j(\vec x,t)~~~.
\end{equation}
Hence the covariant radiation gauge condition is Lorentz boost covariant, although not Lorentz boost invariant.
Analogous arguments apply to $\Psi_0=0$ gauge.  Under the Lorentz boost of Eq. \eqref{eq:boost1}, $\Psi_0 \to \Psi_0 -\vec v \cdot \vec \Psi$, so
$\Psi_0=0$ gauge is not preserved, but it can be restored by a compensating infinitesimal gauge transformation.

Referring now to Eq. \eqref{c10}, we note that the covariant radiation gauge Dirac bracket and the anticommutation relations are invariant
under infinitesimal Rarita-Schwinger gauge transformations, such as that of Eq. \eqref{eq:gaugerestore}, up to a
remainder that is quadratic in the gauge parameter.   Hence
the covariant radiation gauge Dirac bracket and the anticommutation relations following from it are Lorentz invariant, since a finite Lorentz
transformation can be built up from a series of infinitesimal ones.

\section{Compatibility of Lorentz covariance and mode counting in on-shell Rarita-Schwinger particle-photon scattering }
We address finally the question  \cite{witten} of whether leading order on-shell scattering of Rarita-Schwinger fields from an external
electromagnetic field has the requisite relativistic covariance, while preserving the correct counting of massless spin $\frac{3}{2}$
propagation modes.  The operator effective action for this scattering process can be read off from the interaction term in Eq. \eqref{eq:action},
\begin{align}\label{eq:effaction}
S_{\rm eff}(\psi_{\mu},A_{\nu}) = &\int d^4x {\cal L}_{\rm eff}(\psi_{\mu},A_{\nu})~~~,\cr
{\cal L}_{\rm eff}(\psi_{\mu},A_{\nu})=& \frac{1}{2}g\,\overline{\psi}_{\mu} (x)
i\epsilon^{\mu\eta\nu\rho}\gamma_5\gamma_{\eta}A_{\nu}(x) \psi_{\rho}(x)~~~,\cr
\end{align}
where we have suppressed spinor indices as in the text from Eq. \eqref{eq:eqmo} onwards.
Since we are taking the external field $A_{\nu}$ as Abelian, the covariant derivatives in the
equations of motion and constraints are given by
\begin{equation}\label{eq:covdeviv}
D_{\nu}=\partial_{\nu}+gA_{\nu}~~,~~\overleftarrow{D}_{\nu}=\overleftarrow{\partial}_{\nu}-gA_{\nu}~~~.
\end{equation}
To keep the analysis simple, we shall assume that $A_{\nu}(x)$ is of short range, and vanishes for $|\vec x|>R$ for
some radius $R$.
This effective action, the equations of motion of Eqs. \eqref{eq:eqmo} and \eqref{eq:eqmo1a}, and the primary and secondary
constraints following from them, given in Eqs. \eqref{eq:constraint1} and \eqref{eq:constraint2}, are all relativistically covariant.  Hence
to do a covariant calculation, it is only necessary to choose a relativistically covariant gauge fixing, which for convenience we take
as the gauge covariant Lorentz gauge condition
\begin{equation}\label{eq:gaugecovlor}
D^{\mu}\psi_{\mu}=0~~~,
\end{equation}
which is attainable from a generic gauge by the gauge transformation of Eq. \eqref{eq:gaugetrans}, provided that $D^{\mu}D_{\mu}$ is invertible.

Taking the matrix element of Eq. \eqref{eq:effaction} between an incoming Rarita-Schwinger state  of four-momentum $p$, and an outgoing Rarita-Schwinger
state of four momentum $p'$, we get the corresponding scattering amplitude
\begin{equation}\label{eq:amplitude}
{\cal A} = \frac{1}{2}ig\,\int d^4x
 \overline{\psi}_{\mu }(p^{\prime},x)
\epsilon^{\mu\eta\nu\rho}\gamma_5\gamma_{\eta}A_{\nu}(x) \psi_{\rho}(p,x)~~~,
\end{equation}
where $\psi_{\rho}$ and $\overline{\psi}_{\mu}$ are now wave functions, rather than operators, that
obey the Rarita-Schwinger equations of motion in the presence of the external field $A_{\nu}$.

We now introduce source currents for the gauge potential $A_{\nu}$ and the Rarita-Schwinger
wave functions $\psi_{\rho}$ and $\overline{\psi}_{\mu}$, and study their conservation properties.  The source current
to which the gauge potential $A_{\nu}$ couples is  defined by writing the scattering amplitude as
\begin{align}\label{eq:source1}
{\cal A}=&\frac{1}{2}ig\,\int d^4x A_{\nu}(x) J^{\nu}(x)~~~,\cr
J^{\nu}(x)=& \overline{\psi}_{\mu }(p^{\prime},x)
\epsilon^{\mu\eta\nu\rho}\gamma_5\gamma_{\eta} \psi_{\rho}(p,x)~~~.\cr
\end{align}
The  source current for the Rarita-Schwinger field $\overline{\psi}_{\mu }(p^{\prime},x)$ is
defined by writing the scattering amplitude as
\begin{align}\label{eq:source2}
{\cal A}=&\frac{1}{2}ig\,\int d^4x\overline{\psi}_{\mu }(p^{\prime},x){\cal J}^{\mu}(p,x)~~~,\cr
{\cal J}^{\mu}(p,x)=&\epsilon^{\mu\eta\nu\rho}\gamma_5\gamma_{\eta}A_{\nu}(x) \psi_{\rho}(p,x)~~~.\cr
\end{align}
Finally, the source current  for the Rarita-Schwinger field $ \psi_{\rho}(p,x)$ is
defined by writing the scattering amplitude as
\begin{align}\label{eq:source3}
{\cal A}=&\frac{1}{2}ig\,\int d^4x \overline{\cal J}^{\rho}(p^{\prime},x) \psi_{\rho}(p,x)~~~,\cr
\overline{\cal J}^{\rho}(p^{\prime},x)=&\overline{\psi}_{\mu }(p^{\prime},x)\epsilon^{\mu\eta\nu\rho}\gamma_5\gamma_{\eta}A_{\nu}(x) ~~~.\cr
\end{align}

We now show that the three currents that we have just defined are conserved.  For the source current  $J^{\nu}$  for the gauge potential,
we have
\begin{align}\label{eq:cons1}
 \partial_{\nu}J^{\nu}=&
  \overline{\psi}_{\mu }(p^{\prime},x)\overleftarrow{D}_{\nu}
\epsilon^{\mu\eta\nu\rho}\gamma_5\gamma_{\eta} \psi_{\rho}(p,x)\cr
 +& \overline{\psi}_{\mu }(p^{\prime},x)
\epsilon^{\mu\eta\nu\rho}\gamma_5\gamma_{\eta} D_{\nu}\psi_{\rho}(p,x)\cr
=&0~~~,\cr
\end{align}
where the first and second terms on the right vanish by the Rarita-Schwinger equations
for $ \overline{\psi}_{\mu }(p^{\prime},x)$ and $\psi_{\rho}(p,x)$ respectively.
For the source current ${\cal J}^{\mu}(p,x)$ for the spinor $\overline{\psi}_{\mu }(p^{\prime},x)$ , we
have
\begin{align}\label{eq:cons2}
D_{\mu}{\cal J}^{\mu}(p,x)=&\epsilon^{\mu\eta\nu\rho}\gamma_5\gamma_{\eta}\big(\partial_{\mu}A_{\nu}(x)\big) \psi_{\rho}(p,x)\cr
+&\epsilon^{\mu\eta\nu\rho}\gamma_5\gamma_{\eta}A_{\nu}(x) D_{\mu}\psi_{\rho}(p,x)\cr
=&0~~~.\cr
\end{align}
The second term on the right vanishes by the Rarita-Schwinger equation
for  $\psi_{\rho}(p,x)$, while the first term on the right can be rewritten as
\begin{equation}\label{eq:cons2a}
\frac{1}{2}\epsilon^{\mu\eta\nu\rho}\gamma_5\gamma_{\eta}F_{\mu\nu}(x)\psi_{\rho}(p,x)
\end{equation}
and vanishes by the secondary constraint of Eq. \eqref{eq:constraint2}.
Finally, for the source current $\overline{\cal J}^{\rho}(p^{\prime},x)$ for the spinor $ \psi_{\rho}(p,x)$, we have
\begin{align}\label{eq:cons3}
\overline{\cal J}^{\rho}(p^{\prime},x)\overleftarrow{D}_{\rho}=&\overline{\psi}_{\mu }(p^{\prime},x)\epsilon^{\mu\eta\nu\rho}\gamma_5\gamma_{\eta}\big(\partial_{\rho}A_{\nu}(x)\big)\cr
+&\overline{\psi}_{\mu }(p^{\prime},x)\overleftarrow{D}_{\rho}\epsilon^{\mu\eta\nu\rho}\gamma_5\gamma_{\eta}A_{\nu}(x)\cr
=&0~~~.\cr
\end{align}
Again, the second term on the right vanishes by the Rarita-Schwinger equation, while the first term on the right
vanishes by the secondary constraint of Eq. \eqref{eq:constraint2}.

Consider now the following three gauge transformations,
\begin{align}\label{eq:gauge1}
A_{\nu}(x) \to   & A_{\nu}(x) + \partial_{\nu}\Lambda~~~,\cr
\psi_{\rho}(p,x) &\to \psi_{\rho}(p,x)+ D_{\rho} \alpha~~~,\cr
\overline{\psi}_{\mu }(p^{\prime},x) \to & \overline{\psi}_{\mu }(p^{\prime},x) + \overline{\beta}\overleftarrow{D}_{\mu}~~~,\cr
\end{align}
with $\alpha$ and $\beta$ independent spinorial gauge parameters. From Eqs. \eqref{eq:source1}-\eqref{eq:source3},
together with Eqs. \eqref{eq:cons1}-\eqref{eq:cons3},
we find that these transformations all leave the amplitude ${\cal A}$ invariant,
\begin{align}
\delta_{\Lambda} {\cal A}= & \frac{1}{2}ig\,\int d^4x \big(\partial_{\nu}\Lambda\big) J^{\nu}(x)
=-\frac{1}{2}ig\,\int d^4x \Lambda  \partial_{\nu} J^{\nu}(x)=0~~~,\cr
\delta_{\alpha} {\cal A}=&\frac{1}{2}ig\,\int d^4x \overline{\cal J}^{\rho}(p^{\prime},x) D_{\rho} \alpha
=-\frac{1}{2}ig\,\int d^4x \overline{\cal J}^{\rho}(p^{\prime},x)\overleftarrow{D}_{\rho} \alpha=0~~~,\cr
\delta_{\beta} {\cal A}=& \frac{1}{2}ig\,\int d^4x    \overline{\beta}\overleftarrow{D}_{\mu}{\cal J}^{\mu}(p,x)
= -\frac{1}{2}ig\,\int d^4x    \overline{\beta}D_{\mu}{\cal J}^{\mu}(p,x)=0~~~.\cr
\end{align}

 We next must
specify more precisely the structure of the spinor wave functions entering the formula for ${\cal A}$.  Since the
gauge field $A_{\nu}$ is assumed to vanish in the external region $|\vec x|>R$, the Rarita-Schwinger wave functions obey
free field equations in this region.  So for $|\vec x|>>R$ they can be taken asymptotically as plane waves at $t \to \pm \infty$,
\begin{align}\label{eq:extsoln}
\psi_{\mu}(p^{\prime},x)\sim &u_{\mu}(p^{\prime})e^{i p^{\prime} \cdot x}~~,~~t \to +\infty~~~,\cr
\psi_{\rho}(p,x)\sim & u_{\rho}(p) e^{i p \cdot x}~~,~~t \to -\infty          ~~~.\cr
\end{align}
With these boundary conditions, the formula for the amplitude takes the final form
\begin{equation}\label{eq:amplitudefinal}
{\cal A} = \frac{1}{2}ig\,\int d^4x
 \overline{\psi}^{(-)}_{\mu }(p^{\prime},x)
\epsilon^{\mu\eta\nu\rho}\gamma_5\gamma_{\eta}A_{\nu}(x) \psi^{(+)}_{\rho}(p,x)~~~.
\end{equation}
The out state (-) and in state (+) boundary conditions used here are analogs of the
  boundary conditions used in
 the distorted wave Born approximation \cite{dwba}, which the construction of Eq. \eqref{eq:amplitudefinal} resembles.
Equation  \eqref{eq:amplitudefinal} then gives an approximation to the matrix element for
 Rarita-Schwinger scattering by the gauge potential.

 Finally, using the gauge invariances of ${\cal A}$, we can now address the question of counting degrees of freedom.
 We have imposed one gauge fixing constraint on the Rarita-Schwinger wave functions, given in operator form
 by Eq. \eqref{eq:gaugecovlor}, which implies that the wave functions obey analogous conditions of
 vanishing gauge covariant divergence,
 \begin{equation}\label{eq:gaugecovlor1}
 \overline{\psi}_{\mu}(p^{\prime},x)\overleftarrow{D}^{\mu}= D^{\rho}\psi_{\rho}(p,x)=0~~~.
 \end{equation}
 To see what further conditions can be imposed, we must examine consistency with the Rarita-Schwinger equation
 for the operator $\psi_{\mu}$.  This is most easily done using the alternative form of the Rarita-Schwinger
 equation given in Eq. \eqref{eq:a6},
 \begin{equation}\label{eq:alteqmo}
 \gamma^{\rho}(D_{\nu}\psi_{\rho}-D_{\rho}\psi_{\nu})=0~~~.
 \end{equation}
 Applying $D^{\nu}$ from the left to this equation, we get
 \begin{equation}\label{eq:alteqmo1}
 D^{\nu}D_{\nu} \gamma^{\rho}\psi_{\rho}-\gamma_{\rho} [D^{\nu},D^{\rho}] \psi_{\nu} - \gamma^{\rho} D_{\rho} D^{\nu}\psi_{\nu}=0~~~,
 \end{equation}
 which using $[D^{\nu},D^{\rho}]=gF^{\nu\rho}$ becomes
 \begin{equation}\label{eq:alteqmo2}
  D^{\nu}D_{\nu} \gamma^{\rho}\psi_{\rho}-g \gamma_{\rho} F^{\nu \rho}\psi_{\nu} - \gamma^{\rho} D_{\rho} D^{\nu}\psi_{\nu}=0~~~.
 \end{equation}
 When $D^{\nu}\psi_{\nu}=0$, this equation simplifies to
 \begin{equation}\label{eq:alteqmo3}
  D^{\nu}D_{\nu} \gamma^{\rho}\psi_{\rho}=g \gamma_{\rho} F^{\nu \rho}\psi_{\nu} ~~~.
 \end{equation}
 In regions where the gauge field is nonvanishing, the right hand side of this equation is nonzero, and so we cannot impose the additional constraint
 $\gamma^{\rho}\psi_{\rho}=0$.  This is not surprising, since where the fields are nonzero there is {\it already} a second covariant constraint given by
 the secondary constraint of Eq. \eqref{eq:constraint2}, which in terms of the dual field ${\tilde F}^{\nu\rho} = \frac{1}{2} \epsilon^{\nu\rho\alpha\beta}F_{\alpha\beta}$ takes the form
 \begin{equation}\label{eq:dualform}
  0= \gamma_{\rho} {\tilde F}^{\nu \rho}\psi_{\nu} ~~~,
\end{equation}
which differs in form from the right hand side of Eq. \eqref{eq:alteqmo3} when the fields are not self-dual or anti-self-dual.
 However, in the external region where the gauge field vanishes, Eq. \eqref{eq:alteqmo3} becomes
 \begin{equation}\label{eq:alteqmo4}
 \partial^{\nu}\partial_{\nu} \gamma^{\rho}\psi_{\rho}=0~~~,
 \end{equation}
and we can then impose $\gamma^{\rho}\psi_{\rho}=0$ by making a gauge transformation in the external region that
preserves the original constraint $\partial^{\rho}\psi_{\rho}=0$. This is possible, since the condition for the
transformation $\psi_{\rho} \to \psi_{\rho}+ \partial_{\rho}\alpha$ to preserve $\partial^{\rho}\psi_{\rho}=0$
is $\partial^{\rho}\partial_{\rho}\alpha=0$, which is compatible with the form of Eq. \eqref{eq:alteqmo4}.
This establishes that in the external region we can impose the two Lorentz covariant conditions $\partial^{\rho}\psi_{\rho}=0$ and
$\gamma^{\rho}\psi_{\rho}=0$, which together with the primary constraint of Eq. \eqref{eq:constraint1} suffices
to give the correct degree of freedom counting for the incoming and outgoing Rarita-Schwinger wave functions \cite{alvarez}.  Alternatively,
if we are not concerned to maintain manifest Lorentz covariance, we can make a gauge transformation in the external region to
the gauge $\psi_0= \vec \nabla \cdot \vec \psi=0$ used in \cite{freed} to enumerate Rarita-Schwinger degrees of freedom.

Note that if one were to construct the Born approximation amplitude, in which the Rarita-Schwinger wave functions
in the presence of the gauge field are replaced by plane waves in the interior region where the potential is nonzero, the
arguments given above for compatibility of Lorentz covariance with degree of freedom counting would fail.
The reason for this is that the  spinor source
currents would then no longer be conserved, even to zeroth order in the gauge coupling $g$, because the free particle plane wave solutions
do not obey the secondary constraint of Eq. \eqref{eq:constraint2}.  The non-existence of a satisfactory Born approximation for
Rarita-Schwinger photon scattering agrees with restrictions on massless particle three point functions obtained by ``on-shell''
methods \cite{mcgady}.  It is also related to the fact that the extra gauge terms $\partial_{\nu}\Omega^{\mu}$ in
Eq. (5.27) of \cite{freed} for the free Rarita-Schwinger propagator are an obstacle to setting up a Lippmann-Schwinger equation
for Rarita-Schwinger photon scattering, which could be expanded in a perturbation series in $g$.
To establish compatibility, we have had to use an analog
of the distorted wave Born approximation \cite{dwba}, in which the leading approximation to the amplitude is constructed using interacting rather than free fermion wave functions and does not have a gauge-invariant small coupling, $g\to 0$ limit.

The results just obtained will play a role in the analysis of closed Rarita-Schwinger fermion loops with gauge field vertices. We conjecture that when fermion propagators
corresponding to the interacting wave functions are used in the loops, the usual vertex conservation properties including the known chiral anomalies will hold, while permitting changes of
fermionic gauge of the Rarita-Schwinger fermions in the loops.  We plan to pursue this as a further investigation.

\section{Discussion}
To conclude, we see that unlike the massive case, the massless gauged Rarita-Schwinger equation leads to a consistent theory at both the classical
and quantized levels, and these statements are already foreshadowed in the equations of \cite{velo} when taken to the zero mass limit.  Thus,
non-Abelian gauging of Rarita-Schwinger fields can be contemplated as part of the anomaly cancelation mechanism in constructing grand unified
models.

Our analysis invites a number of extensions:

\begin{enumerate}

\item We have derived the path integral formula and  the constrained Hamilton equations of motion in $\Psi_0=0$ gauge, and shown
positivity of the Dirac bracket in covariant radiation gauge. This is reminiscent of what happens in quantizing non-Abelian
gauge theories, where unitarity is manifest in one gauge, and renormalizability in another.  In the gauge theory case,  one has an apparatus for transforming from one gauge to another, but that remains to be worked out for fermionic gauge transformations of the Rarita-Schwinger fields.
The analysis of Sec. 11, where we computed an approximation to Rarita-Schwinger photon scattering and found full fermionic gauge invariance,
is evidence that at least on shell, where the equations of motion are obeyed, different gauges are equivalent.

\item  The analysis of Sec. 11 shows that to get a nonvanishing approximation to Rarita-Schwinger photon scattering, one has to go
beyond the Born approximation to scattering, by using in and out solutions of the scattering equations in the presence of the
photon perturbation.  (We have assumed the existence of these interacting solutions of the Rarita-Schwinger equations,  but this
remains to be studied.) This is a consequence of the fact that the secondary constraint $\omega=0$, although involving the gauge
fields, is zeroth order in the gauge field coupling $g$ because it arose from application of $g^{-1} D_{\mu}$ to the Rarita-Schwinger
equation of motion.  Thus a quantum theory of Rarita-Schwinger fields in a gauge field background necessarily has a non-perturbative
aspect.
\item A possible exception to the non-perturbative behavior of fermion lines is when the $\vec E$ and $\vec B$
gauge fields are random, since if Eq. \eqref{eq:solve1} is replaced by an average, denoted by AV,
\begin{equation}\label{eq:average1}
\langle \Psi_0\rangle_{\rm AV} \simeq \Big\langle  \frac{\vec Q}{(\vec B)^2} \Big\rangle_{\rm AV} \cdot \langle \vec \Psi \rangle_{\rm AV}~~~,
\end{equation}
it becomes
\begin{equation}\label{eq:average2}
\langle \Psi_0\rangle_{\rm AV} \simeq \vec \sigma \cdot \langle \vec \Psi \rangle_{\rm AV}~~~,
\end{equation}
which is compatible with $\langle \Psi_0\rangle_{\rm AV} =\vec \sigma \cdot \langle \vec \Psi \rangle_{\rm AV}=0$, the customary
free Rarita-Schwinger constraints employed in \cite{freed}.  This heuristic observation suggests that Rarita-Schwinger fields coupled to
quantized gauge fields with zero background gauge field  may have a perturbative $g \to 0$ limit.

\item  We have included in our analysis only fermionic constraints.  However, the action exponent in Eq. \eqref{eq:smatrix2} has the form (in $A_0=0$ gauge)
\begin{align}\label{eq:smatrixexp}
\frac{1}{2}
\int d^4x & \Big[-\vec{\Psi}^{\dagger}
\cdot (\vec B + \vec \sigma \times \vec E) (\vec \sigma \cdot \vec B)^{-1} \vec \sigma \cdot \vec D \times \vec {\Psi}\cr
+&\vec {\Psi}^{\dagger} \cdot \vec \sigma \times \vec D (\vec \sigma \cdot \vec B)^{-1}(\vec B + \vec \sigma \times \vec E) \cdot \vec \Psi
+\vec{\Psi} ^{\dagger} \cdot \vec D \times \vec \Psi -\frac{1}{2} \vec{\Psi}^{\dagger} \times \vec \sigma \cdot \partial_0 \vec{\Psi}\Big] \cr
=&\int d^4x ( \partial_0\vec{\Psi} \cdot \vec P  - {\rm bosonic ~constraints})-\int dt  H~~~,\cr
\end{align}
with $H$ the Hamiltonian obtained from the stress-energy tensor.  We have not included the bosonic constraints in our analysis, and we note that they will have fermionic brackets
with the fermionic constraints $\phi_a, \, \chi_a$. (For a discussion of bosonic versus fermionic constraints, see \cite{junker}.)

\item  In quantizing, we assumed that the gauge fields $\vec A$  are time independent, so that $d/dt$ and $\vec D$ commute.  As already noted,
this assumption can be dropped if the gauge fields are treated as dynamical variables, leading to an extension, yet to be
analyzed, of the bracket structure, again involving fermionic brackets.

\item In showing in the Abelian case that there is no superluminal propagation, the inversion of $\vec \sigma \cdot \vec B$ to get
$\Psi_0$ only required $(\vec B)^2 \neq 0$.  In the non-Abelian case, where $\vec B$ is itself a matrix, the conditions for
invertibility are nontrivial and have yet to be analyzed.

\item  In demonstrating positivity of the anticommutator, we made essential use of the condition \hfill\break$\vec \sigma \cdot \vec \Psi=0$.  Deriving this from the covariant radiation
gauge condition $\vec D \cdot \vec \Psi=0$ assumed the invertibility of $\vec \sigma \cdot \vec D$, and attainability of covariant radiation gauge assumed the
invertibility of $(\vec D)^2$.  The conditions for invertibility of these two operators remain to be studied.
(The open space index theorems of Callias \cite{callias}  and
Weinberg \cite{callias} involve $\vec \sigma \cdot \vec D + i\phi$, with $\phi$ a scalar field, and so do not give information about the invertibility of $\vec \sigma \cdot \vec D$.)

\item The fact that when gauge fields are present, the gauge conditions $\Psi_0=0$ and $\vec D \cdot \vec \Psi=0$ are not equivalent, has
consequence for the study of condensates that give rise to a mass for the initially massless Rarita-Schwinger field.  The natural way to
form a condensate involving the  spin $\frac{3}{2}$ field $\psi_{\mu}$ and a spin $\frac{1}{2}$ field $\lambda$ is through a structure of the
form $\overline{\lambda} \gamma^{\mu} \psi_{\mu}$. If   $\Psi_0$ and
$\vec \sigma \cdot \vec \Psi$  were simultaneously zero, this would translate back in the covariant formalism  to $\gamma^{\mu} \psi_{\mu}=0$, forbidding such condensate
formation. Thus, the possibility of dynamical mass generation for the spin $\frac{3}{2}$ field depends on the inequivalence that we have demonstrated
of the gauge conditions $\Psi_0=0$ and $\vec D \cdot \vec \Psi=0$ in nonzero gauge field backgrounds. An alternative way to see that the gauge condition
$\gamma^{\mu}\psi_{\mu}=0$ cannot be imposed when fields are present is to note that its left chiral reduction is $\Psi_0=\vec \sigma \cdot \vec \Psi$, which in
general conflicts with the secondary constraint $\vec \sigma \cdot \vec B \Psi_0=(\vec B+\vec \sigma \times \vec E)\cdot \vec \Psi$.

\end{enumerate}

\section{Acknowledgements}

I wish to thank Edward Witten for  conversations about  gauging Rarita-Schwinger fields, Rarita-Schwinger scattering from photons,
 and the invertibility of $\vec \sigma \cdot \vec D$.  I also wish to acknowledge the  various people who
asked about the status of gauged Rarita-Schwinger fields  when I gave seminars on \cite{adler}.  Following on the initial draft
of this paper, I had a fruitful correspondence with Stanley Deser and Andrew Waldron about gauge invariance and counting degrees
of freedom when invariance of the action is conditional on a constraint.
\appendix

\section{Notational conventions and useful identities}

We follow in general the notational conventions of the book {\it Supergravity} by Freedman and Van Proeyen \cite{freed}.
The metric $\eta_{\mu\nu}$ is $(-,+,+,+)$ and the Dirac gamma matrices $\gamma_{\mu}\,,\gamma^{\mu}$ obey the Clifford algebra
\begin{equation}\label{a1}
\gamma_{\mu}\gamma_{\nu}+\gamma_{\nu}\gamma_{\mu}=2\eta_{\mu\nu}~~~.
\end{equation}
They are given in terms of Pauli matrices
$\sigma_j$ by
\begin{align}\label{a2}
\gamma_0=-\gamma^0=&\left( \begin{array} {cc}
 0&-1  \\
 1&0 \\ \end{array} \right)~~~,\cr
 \gamma_j=\gamma^j=&\left( \begin{array} {cc}
 0&\sigma_j  \\
 \sigma_j&0 \\
\end{array}\right) ~~~,\cr
\gamma_5=i\gamma_0\gamma_1\gamma_2\gamma_3=&\left( \begin{array} {cc}
 1&0  \\
 0&-1 \\
\end{array}\right) ~~~.\cr
\end{align}
We also note that
\begin{equation}\label{a3}
\epsilon_{0123}=-\epsilon^{0123}=1~~~,
\end{equation}
the left chiral projector $P_L$ is given by
\begin{equation}\label{a4}
P_L=\frac{1}{2}(1+\gamma_5)~~~,
\end{equation}
and the  spinor $\overline{\psi}$ is defined in terms of the adjoint spinor $\psi^{\dagger}$ by
\begin{equation}\label{a5}
\overline{\psi}=\psi^{\dagger}i\gamma^0~~~.
\end{equation}

As noted in \cite{freed}, the Rarita-Schwinger equation of motion can be written in a number of equivalent forms.
When ordinary derivatives are replaced by gauge covariant derivatives, these are the vector-spinor equations
\begin{align}\label{eq:a6}
\epsilon^{\mu\eta\nu\rho}\gamma_{\eta}D_{\nu}\psi_{\rho}=&0~~~,\cr
\gamma^{\eta \nu \rho} D_{\nu}\psi_{\rho}=&0~~~,\cr
\gamma^{\rho}(D_{\nu}\psi_{\rho}-D_{\rho}\psi_{\nu})=&0~~~,\cr
\gamma^{\alpha}D_{\alpha} (D_{\sigma} \psi_{\nu}-D_{\nu}\psi_{\sigma})=&\gamma^{\rho}\Big([D_{\rho},D_{\sigma}]\psi_{\nu}
+[D_{\nu},D_{\rho}]\psi_{\sigma}+ [D_{\sigma},D_{\nu}]\psi_{\rho}\Big)~~~,\cr
\end{align}
with only the fourth line, which is quadratic in the covariant derivative,  involving more than just a substitution $\partial_{\nu} \to D_{\nu}$ in the formulas
of \cite{freed}. Using $\gamma_{\eta}\gamma^{\eta \nu \rho}=2 \gamma^{\nu \rho}$, these also imply the spinor equation
$\gamma^{\nu\rho} D_{\nu}\psi_{\rho}=0$.
These formulas play a role in verifying stress-energy tensor conservation, as does the
identity \cite{rosen}
\begin{equation} \label{a7}
0=\epsilon^{\lambda\sigma\mu\nu}(A_{\tau}B_{\lambda}C_{\sigma}D_{\mu}E_{\nu}
+ A_{\nu}B_{\tau}C_{\lambda}D_{\sigma}E_{\mu}+ A_{\mu}B_{\nu}C_{\tau}D_{\lambda}E_{\sigma}
+ A_{\sigma}B_{\mu}C_{\nu}D_{\tau}E_{\lambda}+ A_{\lambda}B_{\sigma}C_{\mu}D_{\nu}E_{\tau})~~~,
\end{equation}
with $A_{\tau},\,B_{\lambda},\,C_{\sigma},\,D_{\mu},\,E_{\nu}$ five arbitrary four vectors. This identity
follows from
\begin{equation}\label{a8}
0=\delta_{\tau}^{\alpha}\epsilon^{\lambda\sigma\mu\nu}+\delta_{\tau}^{\nu}\epsilon^{\alpha\lambda\sigma\mu}
+\delta_
{\tau}^{\mu}\epsilon^{\nu\alpha\lambda\sigma}+\delta_{\tau}^{\sigma}\epsilon^{\mu\nu\alpha\lambda}+\delta_{\tau}^{\lambda}\epsilon^{\sigma\mu\nu\alpha}~~~,
\end{equation}
which is easily verified  by noting that $\lambda,\,\sigma,\,\mu,\,\nu$ must take
distinct values from the set $0,1,2,3$, and that $\tau$ must be equal to one of these values.

The fundamental identity for the Pauli matrices is
\begin{equation}\label{a9}
\sigma_a\sigma_b=\delta_{ab}+i\epsilon_{abc} \sigma_c~~~,
\end{equation}
with $\epsilon_{123}=1$ and with the index $c$ summed.
We repeatedly use the following two identities that can be derived from Eq. \eqref{a9}, for a general three vector $\vec A$ that is proportional to a unit
matrix in the spinor space and so commutes
 with $\vec \sigma$,
\begin{align}\label{a10}
\vec \sigma \times (\vec \sigma \times \vec A)=&-2 \vec A+ i \vec \sigma \times \vec A~~~,\cr
(\vec A \times \vec \sigma) \times \vec \sigma=&-2 \vec A+ i \vec A \times \vec \sigma~~~.\cr
\end{align}
Additional useful identities  are
\begin{align}\label{a11}
\vec \sigma \times \vec \sigma =& 2i\vec \sigma ~~~,\cr
\vec \sigma \,\vec \sigma \cdot \vec A=& \vec A-i\vec \sigma \times \vec A ~~~,\cr
\vec \sigma \cdot \vec A \, \vec \sigma =& \vec A+i\vec \sigma \times \vec A ~~~, \cr
(\vec \sigma \times \vec A) \cdot \vec \sigma=&-2i \vec \sigma \cdot \vec A~~~,\cr
\vec \sigma \cdot (\vec \sigma \times \vec A)=& 2i \vec \sigma \cdot \vec A~~~,\cr
\sigma_a\sigma_b=&2\big(\delta_{ab}-\frac{1}{2}\sigma_b\sigma_a\big)~~~,\cr
\vec B=i\vec A -\vec A \times \vec \sigma  \leftrightarrow & \vec A=\frac{1}{2}(\vec B\times \vec \sigma)~~~.\cr
\end{align}

Gauge field covariant derivatives are
\begin{equation}\label{a12}
D_{\mu}=\partial_{\mu}+gA_{\mu}~~~,
\end{equation}
with the gauge potential $A_{\mu}=A_{\mu}^A t_A $ and the gauge generators $t_A$ anti-self-adjoint,
and with the components $A_{\mu}^A$ self-adjoint.  The non-Abelian generators $t_A$ obey the compact  Lie algebra
\begin{equation}\label{a13}
[t_A,t_B]=f_{ABC}t_C~~~;
\end{equation}
in the Abelian case we replace $t_A$ by $-i$.  In writing field strengths $\vec E$ and $\vec B$ we pull out
an additional factor of $i$ to make them self-adjoint, so that we have the identities
\begin{align}\label{a14}
\vec D \times \vec D =& -i g \vec B~~~,\cr
[\vec D, D_0] =& -i g \vec E~~~.\cr
\end{align}
We will also write a right-acting three-vector covariant derivative as $\overrightarrow  D=\overrightarrow \nabla +g \vec A$, and define a left-acting three-vector covariant derivative as $\overleftarrow D=\overleftarrow \nabla-g\vec A$, so that we
have the integration by parts formulas
\begin{align}\label{a15}
\int d^3x A \overrightarrow  D_{\vec x} B=&-\int d^3x A  \overleftarrow D_{\vec x} B~~~,\cr
\vec{D}_{\vec x} \delta^3(\vec x -\vec y)=&-\delta^3(\vec x -\vec y)\overleftarrow{D}_{\vec y}~~~.\cr
\end{align}
An analogous definition is used for the operators $\vec L$ and $\overleftarrow{L}$ which enter the gauge fixing condition.

At the classical level, variables will be either Grassmann even or odd.  Irrespective of the Grassmann parity of monomials
$A$ and $B$, the adjoint operation is defined by \cite{freed}
\begin{equation}\label{a16}
(AB)^{\dagger}=B^{\dagger}A^{\dagger}~~~.
\end{equation}
For classical brackets, we follow the convention of Henneaux and Teitelboim \cite{teitel},
\begin{equation}\label{a17}
[F,G]_C=\left(\frac{\partial F}{\partial q^i}\frac{\partial G}{\partial p_i}-
\frac{\partial F}{\partial p_i}\frac{\partial G}{\partial q^i}\right)+(-)^{\epsilon_F}
\left(\frac{\partial^L F}{\partial \theta^{\alpha} }\frac{\partial^L G}{\partial \pi_{\alpha}}+
\frac{\partial^L F}{\partial \pi_{\alpha}}\frac{\partial^L G}{\partial \theta^{\alpha}}\right)~~~,
\end{equation}
with $\epsilon_F$ the Grassmann parity of $F$, with $\partial^L$ a Grassmann derivative acting from the left, and with $q^i,\,p_i$ ($\theta^{\alpha},\,\pi_{\alpha}$) canonical coordinates
and momenta of even (odd) Grassmann parity. Using the classical bracket, the Dirac  bracket is constructed  from
the constraints as in Eq. \eqref{eq:dirac1} of the text.   To make the transition to quantum theory, the quantum commutator (anticommutator) is defined to be $i\hbar$ times the corresponding Dirac bracket (with $\hbar=1$ in our notation).   Classical canonical brackets are always denoted, as above, by
a subscript $C$, with a subscript $D$ used for the corresponding Dirac bracket. We use the standard notations $[A,B]=AB-BA$ for the  commutator and $\{A,B\}=AB+BA$ for the anticommutator.

To calculate the Dirac bracket, we  use block inversion of a matrix.  Let
\begin{align}\label{a18}
M=&\left( \begin{array} {cc} A_1  & A_2 \\A_3 &A_4\\ \end{array} \right)~~~,\cr
M^{-1}=&\left( \begin{array} {cc} B_1  & B_2 \\B_3 &B_4\\ \end{array} \right)~~~,\cr
\end{align}
with $A_1,...,A_4$ themselves matrices.  Then when $A_4$ is non-singular, the blocks $B_1,...,B_4$ of $M^{-1}$  are given by
\begin{align}\label{a19}
\Delta\equiv& A_1-A_2 A_4^{-1} A_3 ~~~,\cr
B_1=& \Delta^{-1}~~~,\cr
B_2=&-\Delta^{-1}A_2 A_4^{-1}~~~,\cr
B_3=&-A_4^{-1}A_3 \Delta^{-1}~~~,\cr
B_4=&A_4^{-1}+A_4^{-1} A_3 \Delta^{-1}A_2 A_4^{-1}~~~.\cr
\end{align}
Even though the blocks are noncommutative, Eqs. \eqref{a18} and \eqref{a19} give an inverse that obeys $M^{-1}M=MM^{-1}=1$.

When the constraints $\phi_a$ and $\chi_a$ are combined into an 8 element set of constraints $\kappa_a=\phi_a,\,\kappa_{a+4}=\chi_a,\,a=1,...,4$
then the bracket matrix $S_{ab}(\vec x, \vec y)\equiv [\kappa_a(\vec x),\kappa_b(\vec y)]_C $ can be expressed in terms of the matrix $M_{ab}(\vec x,\vec y)$ of Eq. \eqref{eq:nonvanishbracks} as
\begin{equation}\label{a20}
S(\vec x,\vec y)=\left( \begin{array} {cc}
 0&M(\vec x,\vec y)  \\
 M^T(\vec y,\vec x)&0 \\ \end{array} \right)
 ~~~~,
\end{equation}
where $M_{ab}^T(\vec x, \vec y)=M_{ba}(\vec x, \vec y)$ is the matrix transpose.  Defining the inverse $M^{-1}(\vec x,\vec y)$ that obeys
$\int d^3z M^{-1}(\vec x,\vec z )M(\vec z, \vec y)=\int d^3z M(\vec x,\vec z) M^{-1}(\vec z,\vec y)= \delta^3(\vec x -\vec y)$, it is easy to verify that
\begin{equation}\label{a21}
S^{-1}(\vec x,\vec y)=\left( \begin{array} {cc}
 0&M^{T\,-1}(\vec y,\vec x)  \\
 M^{-1}(\vec x,\vec y)&0 \\ \end{array} \right)
 ~~~~.
\end{equation}

\section{Analysis of the Rarita-Schwinger field in an external Abelian gauge field,
without using the covariant radiation gauge constraint: propagation of the longitudinal mode}

We continue here the analysis begun in Sec. 5, but now without using the covariant radiation gauge constraint
$\vec D \cdot \vec \Psi=0$, to analyze propagation of the longitudinal mode that is eliminated by gauge fixing.
 We must now solve for $C_3^{\uparrow,\,\downarrow}$ starting from Eq.
\eqref{eq:fcomp}  with $C_{1,2}=0$,   so the third line
of Eq. \eqref{eq:fcomp} simplifies to
\begin{align}\label{eq:fcomp3}
0=&(\vec B)^2 \Omega C_3^{\uparrow,\,\downarrow}- K ( Q_3   C_3)^{\uparrow,\,\downarrow}~~~,\cr
Q_3=&B_1E_2-B_2E_1 + B_3 \vec \sigma \cdot (\vec B+i\vec E)-i\vec B\cdot \vec E \sigma_3~~~.\cr
\end{align}
Writing this as
\begin{equation}\label{eq:nmatrixdef}
\left( \begin{array} {c} 0 \\0 \\ \end{array} \right)
=\left( \begin{array} {cc} U_{11} & U_{12} \\U_{21} & U_{22} \\ \end{array} \right)
\left( \begin{array} {c} C_3^{\uparrow} \\ C_3^{\downarrow} \\ \end{array} \right)~~~,
\end{equation}
we find for the matrix elements
\begin{align}\label{eq:matrixelts}
U_{11}=&(\vec B)^2 \Omega-K[B_1E_2-B_2E_1-i(B_1E_1+B_2E_2)+B_3^2]~~~,\cr
U_{22}=&(\vec B)^2 \Omega-K[B_1E_2-B_2E_1+i(B_1E_1+B_2E_2)-B_3^2]~~~,\cr
U_{12}=&-KB_3[B_1+iE_1-i(B_2+iE_2)]~~~,\cr
U_{21}=&-KB_3[B_1+iE_1+i(B_2+iE_2)]~~~.\cr
\end{align}
The equation $0=\rm{det}(U)=U_{11}U_{22}-U_{12}U_{21}$ reduces, after dividing by an overall
factor of $(\vec B)^2$, to
\begin{equation}\label{eq:omegakeq}
0=(\vec B)^2 \Omega^2 - 2 \Omega K (B_1 E_2-B_2 E_1) + K^2 (E_1^2+E_2^2-B_3^2)~~~,
\end{equation}
with the solution
\begin{align}\label{eq:omegaksoln}
\frac{\Omega}{K}=&\frac{X \pm Y^{1/2}}{(\vec B)^2}~~~,\cr
X=& B_1E_2-B_2E_1~~~,\cr
Y=& (B_1E_2-B_2E_1)^2-(\vec B)^2(E_1^2+E_2^2-B_3^2) ~~~.\cr
\end{align}

The analysis of the solutions of Eqs. \eqref{eq:omegakeq} and \eqref{eq:omegaksoln} divides into two cases, according to whether
the roots of Eq. \eqref{eq:omegaksoln} are both real, or both complex. The roots are both complex if
\begin{equation}\label{eq:complex}
(B_1E_2-B_2E_1)^2<(\vec B)^2(E_1^2+E_2^2-B_3^2)~~~,
\end{equation}
which can be rearranged algebraically to the form
\begin{equation}\label{eq:complex1}
[(\vec B)^2-(E_1^2+E_2^2)]B_3^2<(B_1^2+B_2^2)(E_1^2+E_2^2)\cos^2\phi~~~,
\end{equation}
where we have written
\begin{align}\label{eq:phidef}
B_1E_2-B_2E_1=&(B_1^2+B_2^2)^{1/2}(E_1^2+E_2^2)^{1/2} \sin\phi~~~,\cr
B_1E_1+B_2E_2=&(B_1^2+B_2^2)^{1/2}(E_1^2+E_2^2)^{1/2} \cos\phi~~~.\cr
\end{align}
Since the right hand side of Eq. \eqref{eq:complex1} is non-negative, when the
left hand side is negative the inequality is satisfied, and both roots are complex.
Hence a necessary (but not sufficient) condition for both roots to be real is
\begin{equation}\label{eq:realcond}
(\vec B)^2-(E_1^2+E_2^2)>0~~~.
\end{equation}

\subsection{The hyperbolic case: both roots real}

When both roots are real, Eq. \eqref{eq:fcomp3} describes the hyperbolic case of propagating waves.  Introducing
the velocity $V=\Omega/K$, Eq. \eqref{eq:omegakeq} can be written as
\begin{equation}\label{eq:omegakeqnew}
0=(\vec B)^2 V^2-2V(B_1E_2-B_2E_1)+E_1^2+E_2^2-B_3^2~~~,
\end{equation}
which can be rearranged algebraically to the form
\begin{equation}\label{eq:omegakeqnew1}
[(B_1^2+B_2^2)^{1/2}-(E_1^2+E_2^2)^{1/2}]^2+(\vec B)^2(V^2-1)=2(B_1^2+B_2^2)^{1/2}(E_1^2+E_2^2)^{1/2}(V \sin\phi -1)~~~.
\end{equation}
Let us now assume that $V^2>1$, and show that this leads to a contradiction.  When $V^2>1$, the left hand side of
Eq. \eqref{eq:omegakeqnew1} is nonnegative, which implies that $V \sin\phi$ on the right must be nonnegative, and
so can be replaced by its absolute value.  Hence the right hand side of Eq. \eqref{eq:omegakeqnew1} obeys the
inequality
\begin{equation}\label{eq:ineq1}
2(B_1^2+B_2^2)^{1/2}(E_1^2+E_2^2)^{1/2}(V \sin\phi -1)=2(B_1^2+B_2^2)^{1/2}(E_1^2+E_2^2)^{1/2}(|V \sin\phi| -1)
\leq 2 (\vec B)^2 (|V| -1)~~~,
\end{equation}
where we have used Eq. \eqref{eq:realcond}.  But the left hand side of  Eq. \eqref{eq:omegakeqnew1} obeys the inequality
\begin{equation}\label{eq:ineq2}
[(B_1^2+B_2^2)^{1/2}-(E_1^2+E_2^2)^{1/2}]^2+(\vec B)^2(V^2-1)\geq (\vec B)^2(|V|+1) (|V|-1)>2(\vec B)^2(|V|-1)~~~,
\end{equation}
which is a contradiction, since a real number cannot be strictly less than itself.  Hence we must have $V^2 \leq 1$, and
there is no superluminal propagation.

\subsection{The elliptic case: both roots complex}

When both roots are complex, Eq. \eqref{eq:fcomp3} describes the elliptic case in which  there
are no propagating waves; when a propagating wave enters an elliptic region from a hyperbolic one it will be damped to zero
amplitude.  However, in the case of weak damping one can still define a wave velocity and ask what its magnitude is.
When both roots are imaginary, Eq. \eqref{eq:omegaksoln} takes the form
\begin{align}\label{eq:omegaksoln1}
\frac{\Omega}{K}=&\frac{X \pm i(-Y)^{1/2}}{(\vec B)^2}~~~,\cr
X=& B_1E_2-B_2E_1~~~,\cr
-Y=& -(B_1E_2-B_2E_1)^2+(\vec B)^2(E_1^2+E_2^2-B_3^2) ~~~.\cr
\end{align}
Regarding $\Omega$ as real  and the wave number $K$ as complex, the effective propagation velocity has the magnitude
\begin{equation}\label{eq:effvel}
|V_{\rm eff}|=\Big|\frac{\Omega}{K_R}\Big|=\frac{X^2-Y}{(\vec B)^2 |X|}=\frac{E_1^2+E_2^2-B_3^2}{|B_1E_2-B_2E_1|}~~~.
\end{equation}
The condition for weak damping is $-Y<<X^2$, which can be rewritten as
\begin{equation}\label{eq:weakdamp}
(\vec B)^2 (E_1^2+E_2^2-B_3^2)<<2 (B_1E_2-B_2E_1)^2~~~,
\end{equation}
and implies
\begin{equation}\label{eq:weakdamp1}
|V_{\rm eff}|<<\frac{2|B_1E_2-B_2E_1|}{(\vec B)^2}\leq \frac{2 |\Vec E|}{|\vec B|}~~~.
\end{equation}
Hence as long as $2|\vec E|$ is not much larger than $|\vec B|$, which is required by the vacuum stability condition $|\vec E| < |\vec B|$, the damped wave propagation velocity is subluminal.

\section{Construction of the projector $R_{ij}(\vec x,\vec y)$}

Since there are only two $\phi_{a}$ constraints and two $\chi_{a}$ constraints, we index them $a=1,2$ rather than $a=3,4$ as in the text, and use
the invariance of $R_{ij}(\vec x,\vec y)$ under changing the linear combination of the $\chi_{a}$ constraints.
We start from the constraint set
\begin{align}\label{c1}
\phi_1=\vec \sigma \times \vec D \cdot \vec \Psi~~,~~~\chi_1= \vec P \cdot \overleftarrow{D}~~~,\cr
\phi_2=\vec D \cdot \vec \Psi~~,~~~\chi_2= \vec P \cdot \vec \sigma \times \overleftarrow{D}~~~.\cr
\end{align}
For the bracket matrix
\begin{equation}\label{c2}
M_{ab}(\vec x, \vec y)= [\phi_a(\vec x),\chi_b(\vec y)]_C= \left( \begin{array} {cc}
 \hat{\cal A}&\hat{\cal B} \\
 \hat{\cal C}&\hat{\cal D}\\ \end{array} \right)~~~,
\end{equation}
we find the matrix elements
\begin{align}\label{c3}
\hat {\cal A}=& -ig  \vec \sigma \cdot \vec B \delta^3(\vec x-\vec y)    ~~~,\cr
\hat {\cal B}=&  \big(2 (\vec D_{\vec x})^2 + g  \vec \sigma \cdot \vec B \big)  \delta^3(\vec x-\vec y)
=\delta^3(\vec x-\vec y) \big(2  (\overleftarrow{D}_{\vec y})^2  + g  \vec \sigma \cdot \vec B \big)     ~~~,\cr
\hat {\cal C}=&  (\vec D_{\vec x})^2     \delta^3(\vec x-\vec y) =  \delta^3(\vec x-\vec y) (\overleftarrow{D}_{\vec y})^2   ~~~,\cr
\hat {\cal D}=&    ig  \vec \sigma \cdot \vec B \delta^3(\vec x-\vec y)    ~~~.\cr
\end{align}
We write the inverse matrix $M^{-1}(\vec z,\vec w)$ as
\begin{equation}\label{c4}
\left( \begin{array} {cc}
 \hat{\cal F}&\hat{\cal G} \\
 \hat{\cal H}&\hat{\cal I}\\ \end{array} \right)~~~,
 \end{equation}
which obeys
\begin{equation}\label{c5}
\left( \begin{array} {cc}
 \hat{\cal A}&\hat{\cal B} \\
 \hat{\cal C}&\hat{\cal D}\\ \end{array} \right)\left( \begin{array} {cc}
 \hat{\cal F}&\hat{\cal G} \\
 \hat{\cal H}&\hat{\cal I}\\ \end{array} \right)
 =\left( \begin{array} {cc}
 \hat{\cal F}&\hat{\cal G} \\
 \hat{\cal H}&\hat{\cal I}\\ \end{array} \right)\left( \begin{array} {cc}
 \hat{\cal A}&\hat{\cal B} \\
 \hat{\cal C}&\hat{\cal D}\\ \end{array} \right)=
 \left( \begin{array} {cc}
 1&0 \\
 0&1\\ \end{array} \right)~~~.
 \end{equation}
In terms of the inverse matrix,  the projector $R_{ij}(\vec x,\vec w)$ is given by (with internal symmetry indices suppressed)
\begin{align}\label{c6}
R_{ij}(\vec x,\vec w)=& \delta_{ij} \delta^3(\vec x-\vec w) 1\cr
 +& D_{\vec x \,i}\hat{\cal F}(\vec x-\vec w) (\vec \sigma \times \overleftarrow{D}_{\vec w })_j+ D_{\vec x \,i}\hat{\cal G}(\vec x-\vec w) \overleftarrow{D}_{\vec w \,j}\cr
+&(\vec \sigma \times D_{\vec x })_i \hat{\cal H}(\vec x-\vec w)(\vec \sigma \times \overleftarrow{D}_{\vec w })_j
+(\vec \sigma \times D_{\vec x })_i \hat{\cal I}(\vec x-\vec w)\overleftarrow{D}_{\vec w \,j}~~~.\cr
\end{align}
From this expression, we find
\begin{equation}\label{c7}
D_{\vec x\,i}R_{ij}(\vec x,\vec w)=R_{ij}(\vec x,\vec w)\overleftarrow{D}_{\vec w \,j}
=(\vec \sigma \times D_{\vec x })_i R_{ij}(\vec x,\vec w)=R_{ij}(\vec x,\vec w)(\vec \sigma \times \overleftarrow{D}_{\vec w })_j=0~~~.
\end{equation}
In verifying these, it is not necessary to evaluate the inverse matrix; instead, after contracting on the vector index $i$ or $j$ one
expresses the resulting pre- or post- factor in terms of $\hat{\cal A},...,\hat{\cal D}$ and then uses the algebraic relations following
from multiplying out the matrices in Eq. \eqref{c5}.  Finally, contracting
\begin{align}\label{c8}
\vec \sigma \cdot \vec D_{\vec x} \sigma_i=&(D_{\vec x} + i\vec \sigma \times \vec D_{\vec x})_i~~~,\cr
\sigma_j \vec \sigma \cdot \overleftarrow{D}_{\vec w}=&(\overleftarrow{D}_{\vec w}-i \vec \sigma \times \overleftarrow{D}_{\vec w})_j~~~,\cr
\end{align}
with $R_{ij}(\vec x,\vec w)$, we conclude that
\begin{equation}\label{c9}
\sigma_i R_{ij}(\vec x,\vec y) =R_{ij}(\vec x,\vec y) \sigma_j=0~~~,
\end{equation}
when $\vec \sigma \cdot \vec D$ is invertible.

As a consequence of Eqs. \eqref{eq:projectordef} and  \eqref{c7}, $\tilde \Psi_i(\vec x)$ is invariant under the transformations
\begin{align}\label{c10}
\vec \Psi   \to& \vec \Psi + \vec D \epsilon~~~,\cr
\vec \Psi \to & \vec \Psi + \vec \sigma \times \vec D \epsilon~~~.\cr
\end{align}
The first of these implies that the canonical anticommutation relations are invariant under infinitesimal
Rarita-Schwinger gauge transformations starting from covariant radiation gauge.

\end{document}